\documentclass[journal]{IEEEtran}

\usepackage{amsmath}
\usepackage{amsmath,bm}
\usepackage{amsfonts} 
\usepackage{graphicx} 
\usepackage{caption}
\usepackage{subcaption}
\usepackage{array}

\usepackage{multirow}

\usepackage{diagbox}

\usepackage[T1]{fontenc}

\usepackage{bm}
\usepackage{amsfonts,amssymb}

\ifCLASSINFOpdf
\else

\fi

\begin{document}

\title{CIS-Net: A Novel CNN Model for Spatial Image Steganalysis via Cover Image Suppression}
\author{Songtao Wu,
        Sheng-hua Zhong*,~\IEEEmembership{Member,~IEEE,}
        Yan Liu, and Mengyuan Liu
\thanks{Songtao Wu and Sheng-hua Zhong are with the College of Computer Science and Software Engineering, Shenzhen University, Shenzhen, China. Email: \{csstwu, csshzhong\}@szu.edu.cn. Sheng-hua Zhong is the corresponding author of this paper.}
\thanks{Yan Liu is with the Department of Computing, The Hong Kong Polytechnic University, Hong Kong SAR, China. Email: csyliu@comp.polyu.edu.hk.}
\thanks{Mengyuan Liu now is with the Tencent Research, Shenzhen, China. Email: nkliuyifang@gmail.com.} }

\markboth{Submission to IEEE Transactions on XXX }%
{Paper name}
\maketitle

\begin{abstract}
Image steganalysis is a special binary classification problem that aims to classify natural cover images and suspected stego images which are the results of embedding very weak secret message signals into covers. How to effectively suppress cover image content and thus make the classification of cover images and stego images easier is the key of this task. Recent researches show that Convolutional Neural Networks (CNN) are very effective to detect steganography by learning discriminative features between cover images and their stegos. Several deep CNN models have been proposed via incorporating domain knowledge of image steganography/steganalysis into the design of the network and achieve state of the art performance on standard database. Following such direction, we propose a novel model called Cover Image Suppression Network (CIS-Net), which improves the performance of spatial image steganalysis by suppressing cover image content as much as possible in model learning. Two novel layers, the Single-value Truncation Layer (STL) and Sub-linear Pooling Layer (SPL), are proposed in this work. Specifically, STL truncates input values into a same threshold when they are out of a predefined interval. Theoretically, we have proved that STL can reduce the variance of input feature map without deteriorating useful information. For SPL, it utilizes sub-linear power function to suppress large valued elements introduced by cover image contents and aggregates weak embedded signals via average pooling. Extensive experiments demonstrate the proposed network equipped with STL and SPL achieves better performance than rich model classifiers and existing CNN models on challenging steganographic algorithms.

\end{abstract}


\section{Introduction}
As the development of social media, huge amount of digital images are uploaded to internet every day. The proliferation of digital images on the internet provide intended users accessible media for criminal purpose easily. Image steganography, the science and art to hide secret messages into images by slightly modifying their pixels/coefficients, is one of key methods for covert communication with digital images [1-8]. As a counterpart of image steganography, image steganalysis is the technique to reveal the presence of secret message in a digital image [9-14]. Because of its importance of information security, image steganalysis has been developed greatly in recent years [15-16].

Steganalysis for natural images in spatial domain proves to be a difficult task. Modern steganographic algorithms [2-8] embed secret messages into cover images by modifying each pixel of a cover image with a very small amplitude ($\pm1$). Furthermore, the STC embedding scheme [17] enables steganographers to change pixels located in those complex, cluttered or noisy regions, which are difficult to be accurately modeled by statistical methods. Previous researches [11,18] indicated that it is difficult to learn discriminative features to classify cover images and their stegos when they are fed into a binary classifier directly. Consequently, designing effective features and learning methods that both preserve the embedded message and suppress the cover image content are essential for image steganalysis. In [11], the author proposed to model the differences between adjacent pixels rather than original values for feature extraction. This operation actually suppresses the cover image content by removing their low frequency components and thus obtain a great detection improvement to LSB matching steganography. Extending this approach, Fridrich in [12-13] proposed the Spatial Rich Model (SRM) method that uses thirty linear and nonlinear high-pass filters to extract noise residuals from input covers and stegos. With paired training method, i.e. cover images and their stegos are input to the classifier simultaneously, SRM learns a subset of features most sensitive to message embedding for steganalysis. Same idea is also used in Projection-SRM (PSRM) [14], which utilizes many different random vectors to project cover images and their stegos into low dimensions, in order to highlight hidden message and suppress cover image content as much as possible. Although many efforts have been put to design features for image steganalysis, it is still hard to detect steganography accurately and move them into real applications [19].

The development on deep convolutional neural network has opened a new gate for image steganalysis. Recent progresses of CNN on image related tasks [20-24] have demonstrated its powerful ability to describe the distribution of natural images. This ability, however, can be used to model statistical differences between natural cover images and unnatural stego images. Additionally, CNN deep architecture with convolution, pooling and nonlinear mapping provides steganalyzers larger space to extract more effective features than hand-crafted ones [25]. For these reasons, many CNN models have been proposed for image steganalysis in recent years. Tan in [26] firstly proposed a stacked auto-encoder network to detect steganography. Results in their paper showed that a CNN model suitable for image recognition may not be applicable to image steganalysis directly. In [27], Qian et al. proposed a new neural network equipped with fixed KV high-pass filtering layer and Gaussian action function to detect steganography. This is the first deep learning based steganalyzer that uses domain knowledge of steganalysis, i.e. design special layers to suppress cover image content. Along this direction, Xu [28] proposed a new CNN model which contains absolute value layer, batch normalization layer, TanH layer and $1 \times 1$ convolutional layer. The purpose of these designs is to make the network specialized to the image steganalysis task and prevent overfitting. Following the Xu-network, Li in [29] extended the model with diverse activation modules and made the network achieve much better performance. Wu et al. in [30-31] proposed to use residual connections in a steganalytic network and obtained low detection error rates when cover images and their stegos are trained and tested in pairs. Different from previous models that only use one fixed kernel to suppress cover content, Ye et al. in [32] utilized all thirty SRM high-pass filters in first layer of their network. Additionally, the author proposed a linear truncation layer and a module to incorporate the selection channel information into the design, obtaining significant performance improvement to classic SRM steganalyzer. Recently, Wang in [33] and Boroumand in [34] proposed two clean end-to-end architectures to detect steganography with residual learning. Without any pre-calculated convolutional layers, both networks can automatically learn out high-pass filters to suppress cover image content.

In parallel with steganalysis in spatial domain, deep learning based steganalyzer for JPEG images is also developed in recent years. Based on Xu-network, Chen in [35] and Zeng in [36] replaced the KV kernel with JPEG-phase-aware filters, such as DCT basis patterns and 2D Gabor filters, for suppressing JPEG image content. In [37], Xu proposed a deep residual network with fixed DCT preprocessing filters for JPEG image steganalysis. Use a same end-to-end model to the spatial domain case, Boroumand [34] trained the network with JPEG images and achieved the state of the art performances on the BOSS database. In summary, either the model with predefined or automatically learned kernels, how to effectively suppress cover image content without destroying the existence of embedded message is central for spatial and compressed domain steganalysis with deep neural networks.

Although many efforts have been put to incorporate the domain knowledge of steganalysis into the design of CNN model, how to effectively suppress cover image content is not fully explored. Along this research direction, we propose two novel layers, Single-value Truncation Layer (STL) and Sublinear Pooling Layer (SPL), for cover image content reduction. For STL, it truncates input data into a predefined interval using a same threshold, which is different from a general truncation linear layer that rounds off a large/small value into two different (positive/negative) thresholds. Intuitively, STL can reduce the variance introduced by out-interval elements whose values are truncated with two different thresholds. The assumption is supported by mathematical analysis based on the distribution of natural image pixels. For SPL, it uses the sublinear power function, a function whose power factor is smaller than 1, for cover image content suppression. To avoid destroying the embedded message, SPL aggregates the feature map with average pooling before sublinear suppression. By unifying STL and SPL in a single model, a novel neural network called Cover Image Suppression Network (CIS-Net) is proposed in this paper. Experiments on some challenging steganographic algorithms have demonstrated the superiority of CIS-Net over classic SRMs and existing CNN based steganalyzers. Based on proposed network, we also explore the possibility that a well-learned CNN model is able to roughly estimate embedding probability map of given steganographic algorithm.

The rest of the paper is organized as follows. In section II, we introduce the proposed network for image steganalysis in details. The proposed STL and SPL are described and analyzed in this section.  In section III, we conduct several experiments on standard database to demonstrate the effectiveness of proposed network over existing hand-crafted methods and deep learning based method. In the same section, we use the Classification Activation Map (CAM) [38] technique to draw attentional maps learned by CIS-Net for different steganographic algorithms and compare them with ground truth message embedding probability maps. The paper is finally closed with the conclusion in section IV.

\begin{figure*}[t]
   \centering
   \includegraphics[height=8.8cm, width=15cm]{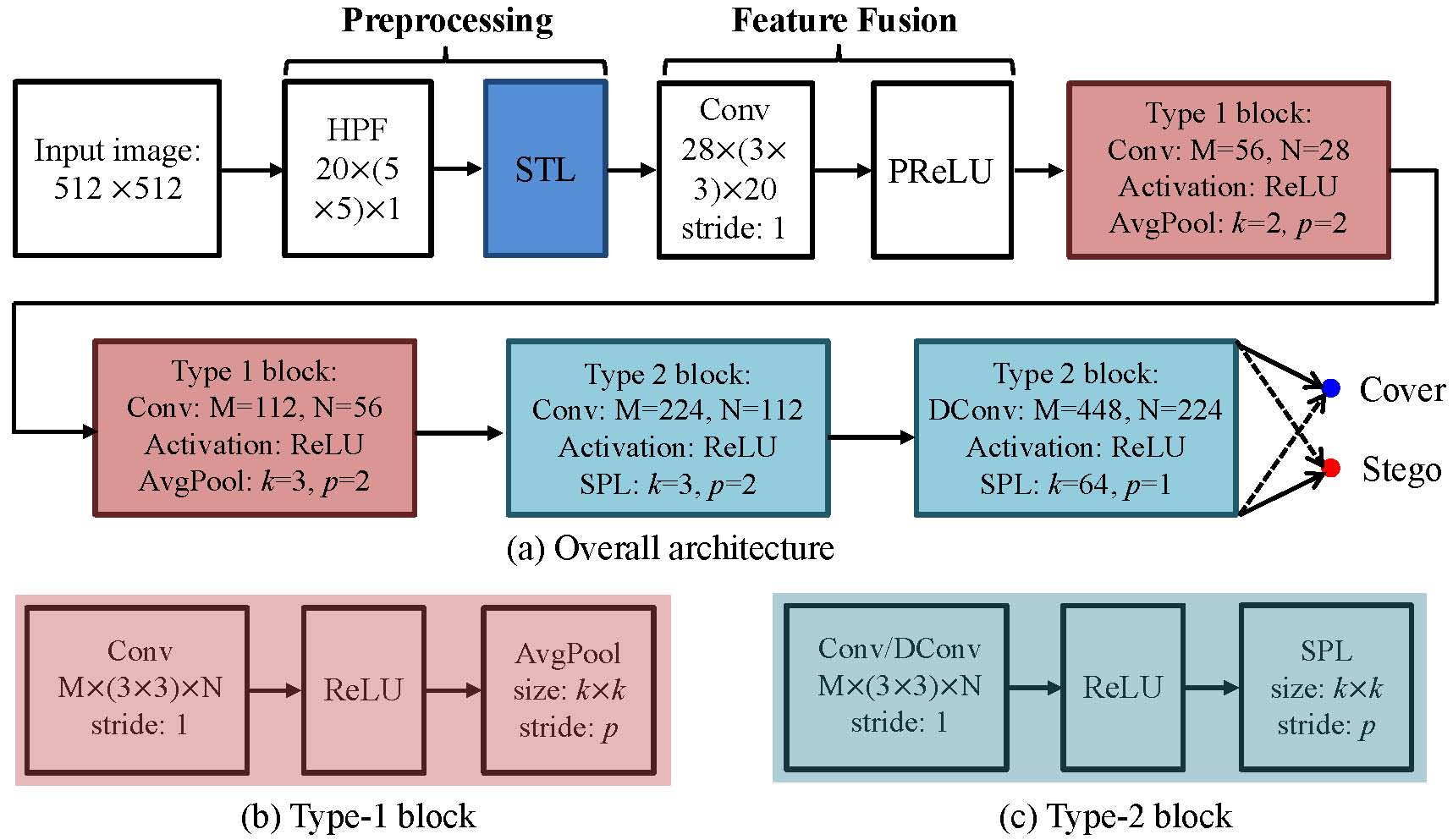}
      \caption{The proposed CIS-Net model for image steganalysis. The whole architecture consists of a preprocessing block, a feature fusion block, two Type-1 blocks and two Type-2 blocks.The preprocessing block suppresses cover image content by extracting high frequency components and using a single-valued truncation layer. The feature fusion block combines different preprocessed information for the following processing. Two Type-1 and Type-2 blocks learn discriminative features for image steganalysis through the suppression of cover image content and the aggregation of embedded message signal.}
\end{figure*}

\section{Proposed Network}
In this section, we introduce the proposed CIS-Net model for image steganalysis. Firstly, the overall architecture of CIS-Net is described in details. Then, we introduce the proposed single-value truncation and sublinear pooling, and explain their rationality for image steganalysis based on theoretical analysis and experiments. 

\subsection{Overall Architecture}
As illustrated in Fig.1, the proposed network contains a preprocessing block, a feature fusion block, two Type-1 blocks and two Type-2 blocks. These building blocks are described in details as follows:

\textbf{Preprocessing block}: The block contains several $5\times 5$ High Pass Filters (HPF) and a STL to preprocess input images. It is noted that image steganalysis is to classify cover images and stego images which are results of adding cover images with very weak high frequency message signal, thus to preprocess input images to make the classification easy is necessary. Specifically, we follow Ye-network's [32] design that use several SRM high pass filters to remove low frequency components and a truncation layer to further filter out those large elements in the cover image. However, there are two main differences between the proposed network and Ye-network. For the first, we refine SRM filters and only select twenty of them for high frequency components extraction. Among all thirty SRM filters, we find the 4-th order HPFs are not beneficial for the effectiveness of our model and they are discarded in the design. The selected high pass filters are shown in Fig.2. For the second, rather than use a traditional truncation method, we propose a new single-valued truncation layer to filter out large elements of cover images. The main advantage of the single-valued truncation is that it can reduce the dynamic range of cover image content compared to traditional truncation method, without destroying preserved information. Details about the method can be found in part B of this section.

\textbf{Feature fusion block}: The block bridges image preprocessing block and the following feature learning blocks. Several $3 \times 3$ convolutional layers in the block fuse different high frequency components extracted from the preprocessed block together and augment features into higher dimensions. Instead of using popular ReLU activation layer, we use a Parametric ReLU (PReLU) after the convolutional layer since it allows information in the negative region pass through the layer, which avoids information loss caused by the ReLU layer.

\textbf{Type-1 block}: Each block uses a unit containing a convolutional layer, a ReLU activation layer and an average pooling layer to extract discriminative feature for image steganalysis. The design is motivated by VGG-net [39] which recursively use $3 \times 3$ convolutional kernels and pooling layers in the network. In the block, there are no batch normalization layers since they may make training be unstable when the mean and variance are not accurately estimated [25].

\begin{figure}[t]
   \centering
   \begin{subfigure}{0.04\textwidth}
     \centering
     \includegraphics[height=0.7cm, width=0.7cm]{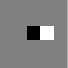}
   \end{subfigure}
   \begin{subfigure}{0.045\textwidth}
     \centering
     \includegraphics[height=0.7cm, width=0.7cm]{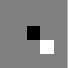}
   \end{subfigure}
   \begin{subfigure}{0.04\textwidth}
     \centering
     \includegraphics[height=0.7cm, width=0.7cm]{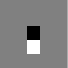}
   \end{subfigure}
   \begin{subfigure}{0.04\textwidth}
     \centering
     \includegraphics[height=0.7cm, width=0.7cm]{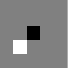}
   \end{subfigure}
   \begin{subfigure}{0.04\textwidth}
     \centering
     \includegraphics[height=0.7cm, width=0.7cm]{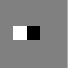}
   \end{subfigure}
   \begin{subfigure}{0.04\textwidth}
     \centering
     \includegraphics[height=0.7cm, width=0.7cm]{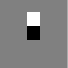}
   \end{subfigure}
   \begin{subfigure}{0.04\textwidth}
     \centering
     \includegraphics[height=0.7cm, width=0.7cm]{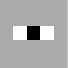}
   \end{subfigure}
   \begin{subfigure}{0.04\textwidth}
     \centering
     \includegraphics[height=0.7cm, width=0.7cm]{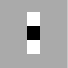}
   \end{subfigure}
   \begin{subfigure}{0.04\textwidth}
     \centering
     \includegraphics[height=0.7cm, width=0.7cm]{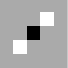}
   \end{subfigure}
   \begin{subfigure}{0.04\textwidth}
     \centering
     \includegraphics[height=0.7cm, width=0.7cm]{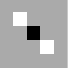}
   \end{subfigure}
   \caption*{(a) Second order and third order high pass filters}

    \begin{subfigure}{0.04\textwidth}
     \centering
     \includegraphics[height=0.7cm, width=0.7cm]{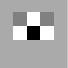}
   \end{subfigure}
   \begin{subfigure}{0.04\textwidth}
     \centering
     \includegraphics[height=0.7cm, width=0.7cm]{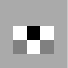}
   \end{subfigure}
   \begin{subfigure}{0.04\textwidth}
     \centering
     \includegraphics[height=0.7cm, width=0.7cm]{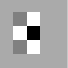}
   \end{subfigure}
   \begin{subfigure}{0.04\textwidth}
     \centering
     \includegraphics[height=0.7cm, width=0.7cm]{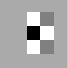}
   \end{subfigure}
   \begin{subfigure}{0.04\textwidth}
     \centering
     \includegraphics[height=0.7cm, width=0.7cm]{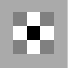}
   \end{subfigure}
   \begin{subfigure}{0.04\textwidth}
     \centering
     \includegraphics[height=0.7cm, width=0.7cm]{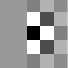}
   \end{subfigure}
   \begin{subfigure}{0.04\textwidth}
     \centering
     \includegraphics[height=0.7cm, width=0.7cm]{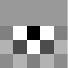}
   \end{subfigure}
   \begin{subfigure}{0.04\textwidth}
     \centering
     \includegraphics[height=0.7cm, width=0.7cm]{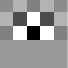}
   \end{subfigure}
   \begin{subfigure}{0.04\textwidth}
     \centering
     \includegraphics[height=0.7cm, width=0.7cm]{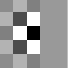}
   \end{subfigure}
   \begin{subfigure}{0.04\textwidth}
     \centering
     \includegraphics[height=0.7cm, width=0.7cm]{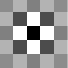}
   \end{subfigure}
   \caption*{(b) KB, KV filters and their variations}
   \caption{Twenty high pass kernels selected from SRM filters as the HPF layers in the proposed CIS-Net. }
\end{figure}

\textbf{Type-2 block}: Each block consists of a $3 \times 3$ convolutional layer, a ReLU activation layer and a sublinear pooling layer. With the help of the proposed sublinear pooling layer (details are introduced in part C of this section), the Type-2 block learns discriminative features from input feature maps by simultaneously aggregating embedded message signal and suppressing cover image content. In order to summarize message information across the whole stego image, we force that the second Type-2 block to uses a large kernel sublinear pooling (kernel size is $64 \times 64$). In addition, the dilated convolution [40] is utilized in this block in order to extract long range correlation from input features.

To summarize, the proposed CIS-Net uses a series of methods to suppress cover image content and preserve embedded message in the network design. All the number of convolutional kernels in the network are optimized based on the task.

\subsection{Single-valued Truncation}
Truncating data into predefined interval proves to be useful either for traditional hand-crafted feature based steganalysis [11-13] or deep learning based steganalysis [25,32]. Compared with cover image content, signal introduced by embedded message is usually low-amplitude. Thus, truncation can filter out cover image content whose elements are usually large amplitude without destroying secret message greatly. In addition, truncation can reduce the dynamic range of input feature map, making the modeling of data's distribution easier [12]. In this subsection, we propose an effective data truncation method for image steganalysis. The method is featured that it preserves the signal of embedded message while reduces the variance of truncated elements.
\subsubsection{Motivation}
in [12,32], truncation is defined as the following equation:
\begin{eqnarray}
  Trunc(x) =
   \left\{
   \begin{array}{lll}
      -T, \ x < -T \\
      x, \ \  -T \leq x \leq T \\
      T, \ \  x > T
   \end{array}
   \right.
\end{eqnarray}
where $T$ is a predefined positive threshold. The $Trunc(\cdot)$ function preserves elements in the interval $[-T, T]$ while maps all other elements into two different values (we call it \textit{bi-valued truncation} in this paper): $T$ for those larger than the predefined positive threshold and $-T$ for those smaller than the negative threshold. Generally, elements of high pass filtered images are symmetrically distributed across zero [41], thus the mean of feature map is zero. Based on such conclusion and Eq.(1), we can write the variance of feature map after bi-valued truncation into three parts:
\begin{equation}
  \sigma_{b}^{2} = \int_{-\infty}^{-T} (-T)^{2} p(x)dx + \int_{-T}^{T} x^{2} p(x)dx + \int_{T}^{+\infty} T^{2} p(x)dx
\end{equation}
where $\sigma_{b}^{2}$ is the element variance after bi-valued truncation, $p(x)$ denotes the probability distribution of the element $x$ after high pass filtering. In Eq.(2), the first term and third term are introduced by two truncated thresholds, while the second term is introduced by preserved elements in $[-T, T]$. Nevertheless, two truncated values $T$ and $-T$ do not provide any useful information for the classification of cover images and stego images, but they increase the variance of feature map by adding the first term and third term in Eq.(2). To reduce the influence of such artificially introduced terms, we propose a novel truncation method called single-valued truncation which is defined as:
\begin{eqnarray}
  STL(x) =
   \left\{
   \begin{array}{lll}
      T, \ |x| > T \\
      x, \ \  -T \leq x \leq T \\
   \end{array}
   \right.
\end{eqnarray}
The main difference between single-valued truncation and bi-valued truncation is that all elements out of the predefined interval $[-T, T]$ are mapped to a same threshold $T$. The variance of feature map element after single-value truncation can be written as:
\begin{equation}
 \begin{aligned}
  \sigma_{s}^{2} & = \int_{-\infty}^{-T} (T-\mu_{s})^{2} p(x)dx + \int_{-T}^{T} (x-\mu_{s})^{2} p(x)dx \\
  & + \int_{T}^{+\infty} (T-\mu_{s})^{2} p(x)dx
 \end{aligned}
\end{equation}
where $\sigma_{s}^{2}$ and $\mu_{s}$ represent the element variance and mean after single-valued truncation respectively. For a symmetrically distributed function $p(x)$, it can easily validate that $\mu_{s}$ is a positive value which is smaller than $T$. Intuitively, we can conclude that the first term and third term in Eq.(4) are decreased compared to two terms in Eq.(2). In the following part, we give a strict proof that $\sigma_{s}^{2}$ is always smaller than $\sigma_{b}^{2}$ for natural images processed by high-pass filters.

\subsubsection{Theoretic Analysis}
In the formulation, we follow the conclusion of previous researches that each pixel in natural images processed by mean-0 highpass filters follows the ``generalized Laplace distribution'' [41-42]:
\begin{equation}
 p(x) = \frac{1}{Z}e^{-\left|\frac{x}{s}\right|^{\alpha}}
\end{equation}
where $\alpha$ and $s$ are two parameters of the distribution, $Z$ is the normalization constant to make the integral of $p(x)$ be 1. For bi-valued truncation, its variance $\sigma_{b}^{2}$ of Eq.(2) can be written as the following formula based on the above distribution:
\begin{equation}
 \begin{aligned}
   \sigma^{2}_{b} = 2\int^{\infty}_{T}\frac{T^{2}}{Z}e^{-\left|\frac{x}{s}\right|^{\alpha}}dx + \int_{-T}^{T}\frac{x^2}{Z}e^{-\left|\frac{x}{s}\right|^{\alpha}}dx
 \end{aligned}
\end{equation}
Based on the symmetrical property of Eq.(5), the mean value of feature map element $\mu_{s}$ for single-valued truncation can be obtained:

\begin{small}
\begin{equation}
 \begin{aligned}
  \mu_{s} & = \int_{-\infty}^{-T} \frac{T}{Z}e^{-\left|\frac{x}{s}\right|^{\alpha}}dx+\int_{-T}^{T} \frac{x}{Z}e^{-\left|\frac{x}{s}\right|^{\alpha}}dx
   +\int_{T}^{\infty} \frac{T}{Z}e^{-\left|\frac{x}{s}\right|^{\alpha}}dx \\ 
  & = 2\int_{T}^{\infty} \frac{T}{Z}e^{-\left|\frac{x}{s}\right|^{\alpha}}dx
 \end{aligned}
\end{equation}
\end{small}
and the variance of Eq.(4) is calculated as:

\begin{figure}[t]
   \centering
   \begin{subfigure}{.24\textwidth}
     \includegraphics[height=3.6cm, width=4.5cm]{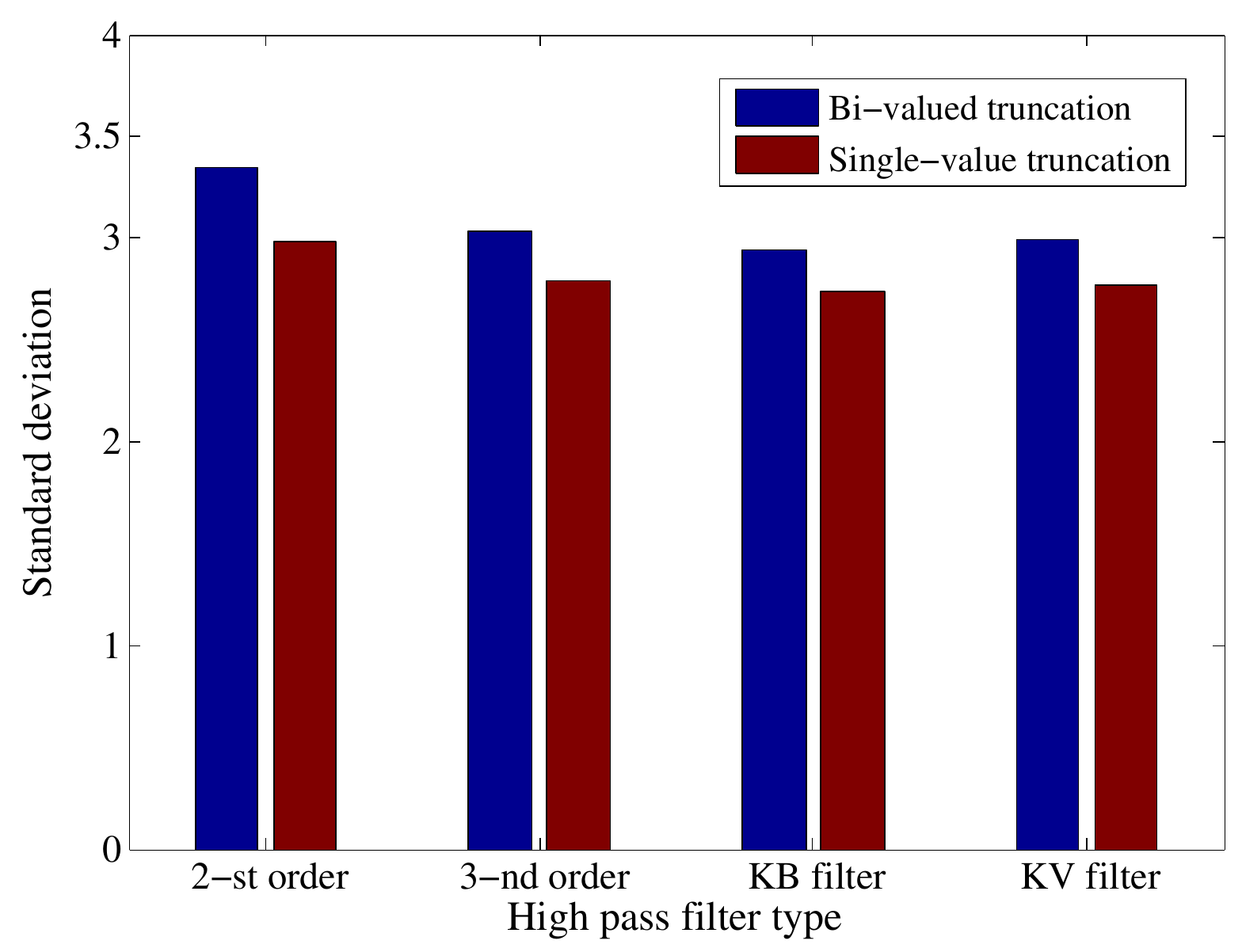}
     \caption{Standard deviations}
   \end{subfigure}
   \begin{subfigure}{.24\textwidth}
     \includegraphics[height=3.6cm, width=4.5cm]{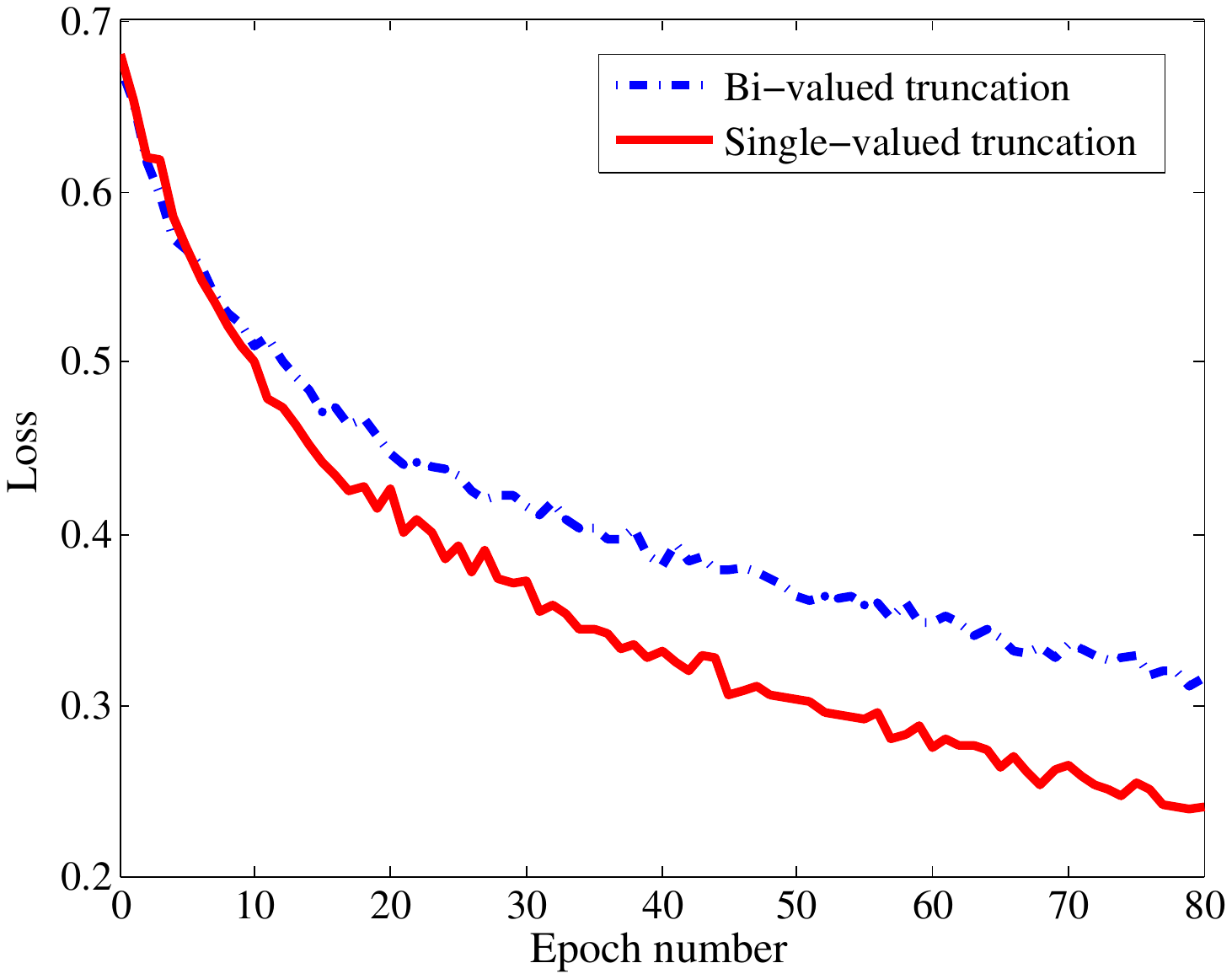}
     \caption{Training loss curves}
   \end{subfigure}
   \caption{Standard deviation and training loss curves. The standard deviation is calculated by 100 randomly selected cover image which are processed by two truncation methods. For training loss curves, we use the proposed architecture except for the different truncation layers to detect S-UNIWARD steganography at 0.4 bpp. Both bi-value truncation and single-valued truncation use the same threshold, i.e. $T=5$. }
\end{figure}

\begin{small}
\begin{equation}
 \begin{aligned}
   \sigma^{2}_{s} = 2\int_{T}^{\infty}\frac{(T-\mu_{s})^{2}}{Z}e^{-\left|\frac{x}{s}\right|^{\alpha}}dx + \int_{-T}^{T}\frac{(x-\mu_{s})^2}{Z}e^{-\left|\frac{x}{s}\right|^{\alpha}}dx
 \end{aligned}
\end{equation}
\end{small}
After some mathematical operations which is illustrated in Appendix, the variance $\sigma^{2}_{s}$ can be rewritten as:
\begin{equation}
 \begin{aligned}
   \sigma^{2}_{s}  = \int_{-T}^{T}\frac{x^2}{Z}e^{-\left|\frac{x}{s}\right|^{\alpha}}dx
   + 2\int_{T}^{\infty}\frac{T^{2}}{Z}e^{-\left|\frac{x}{s}\right|^{\alpha}}dx  - \mu_{s}^{2}
 \end{aligned}
\end{equation}
Based on Eq.(6) and Eq.(9), the difference between $\sigma^{2}_{s}$ and $\sigma^{2}_{b}$ is:
\begin{equation}
 \begin{aligned}
  \sigma^{2}_{s}-\sigma^{2}_{b} & = -\mu_{s}^{2} < 0
 \end{aligned}
\end{equation}
Eq.(10) indicates that, for any positive threshold ($T>0$), the variance of element in the feature map after the single-value truncation is always smaller than the variance after the bi-value truncation. The result demonstrates that the proposed single-valued truncation reduces the variance of traditional bi-valued truncation without deteriorating those preserved elements in the interval $[-T, T]$.

Except for the theoretic analysis, the bi-valued truncation and single-valued truncation are compared on real images and steganographic algorithms. We randomly select 100 cover images from BOSSbase 1.01 and use two truncation methods to process them. In addition, the proposed model with bi-valued truncation and single-valued truncation are learned to detect Spatial UNIversal WAvelet Relative Distortion (S-UNIWARD) steganography [6] at payload 0.4 bpp. Fig.3(a) shows the single-valued truncation decreases standard deviations across all high pass filters, while Fig.3(b) demonstrate that the model with  single-valued truncation converges much faster than the model with bi-valued truncation.


\subsection{Sublinear Pooling}
For deep learning based image steganalysis, it is important to aggregate weak signal of embedded message. Previous researches [27-32] shows that average pooling is better than max pooling for image steganalysis since it can effectively merge embedded message signal in a local region and strengthen embedded message signal across the whole stego image. However, average pooling method only focuses on the aggregation of embedded message but do not take the suppression of cover image content into account. To overcome this limitation, we propose a novel pooling method called sublinear pooling which is depicted by Fig.4. The proposed sublinear pooling unifies both embedded message aggregation and cover content suppression in one block. Mathematically, the sublinear pooling is defined as follows:
\begin{equation}
 SPL(\bm{x};\gamma_{1}, \gamma_{2}) = f_{2}\left(avg\_pool\left(f_{1}(\bm{x};\gamma_{1})\right);\gamma_{2}\right)
\end{equation}
where $avg\_pool$ denote the average pooling, both $f_{1}$ and $f_{2}$ are element-wise sublinear power function $f$ parameterized by a value $\gamma$:
\begin{equation}
 f(\bm{x};\gamma) = |\bm{x}|^{\gamma}\circ sgn(\bm{x}), \ 0< \gamma \leq 1
\end{equation}
where $|\cdot|$, $sgn(\cdot)$ and $\circ$ represent the element-wise absolute value, sign function and multiplication. The element-wise function used in sublinear pooling is motivated by the ``generalized $\alpha$-pooling'' which is proposed in [43]. This design is easy to be implemented in code and optimized by back-propagation algorithm. Additionally, power function satisfies the following inequality:
\begin{equation}
   |x|^{\gamma} \leq |x| \ \ for \ |x| \geq 1, \ 0 < \gamma \leq 1
\end{equation}
This property can be used for cover image suppression when the parameter is set accordingly.

The main difference between the proposed pooling and average pooling is that our method adds sub-linear power functions before (pre-sublinear) and after (post-sublinear) an ordinary average pooling layer. This change brings two advantages for image steganalysis:
\begin{itemize}
  \item Sublinear pooling suppress cover image content adaptively. In the proposed pooling method, sublinear power functions decrease values of elements with large amplitudes. The larger the element's amplitude is, the more  the sublinear function reduces the element. Since large valued elements are mainly generated by cover image, sub-linear pooling decreases their amplitudes and thus suppress cover image content;
  \item Sublinear pooling aggregates embedded message signal effectively. In the middle of two sublinear power functions, average pooling in the proposed method merges message signals from input feature maps that cover image contents are firstly suppressed by pre-sublinear. Then, the post-sublinear further reduces the cover image content in features maps whose embedded message signals are already augmented by the averaging pooling. Such ``suppression-aggregation-suppression'' is more effective than a single average pooling for image steganalysis.
\end{itemize}

\begin{figure}[t]
   \centering
   \includegraphics[height=2.0cm, width=7cm]{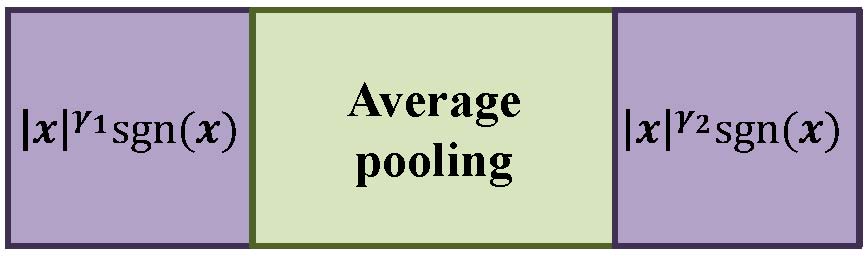}
      \caption{Schematic illustration of proposed sublinear pooling layer. The layer adds the element-wise power function before and after an average pooling layer. To make the power function to be sublinear, $\gamma_{1}$ and $\gamma_{2}$ in the layer should be positive and smaller than 1.   }
\end{figure}

{\setlength{\abovecaptionskip}{2pt}
 \setlength{\belowcaptionskip}{-2pt}
\begin{table}[t]
  \centering
  \renewcommand\arraystretch{1.2}
  \caption{Detection error rates of proposed model with averaging pooling and sublinear pooling for training, testing and their differences on S-UNIWARD steganography at 0.4 bpp. SPL's parameters, $\gamma_{1}$ and $\gamma_{2}$, are set to 0.9 and 0.9 according to the experimental results. }
  \resizebox{8.0cm}{!} {
  \begin{tabular}{| c | c | c | c |}
   \hline
  \textbf{Pooling method} & Training & Testing & Difference \\ \hline
  Average pooling & 6.72\% & 15.11\% & 8.39\% \\ \hline
  Proposed SPL & 10.76\% & 14.82\% & 4.06\% \\ \hline

  \end{tabular}
  }
\end{table}}

To validate the effectiveness of proposed pooling method, we compare training and testing detection error rates of the proposed architecture on S-UNIWARD steganography at 0.4 bpp in two different cases.  In the first case, both Type-1 and Type-2 blocks use averaging pooling method in the whole architecture. In the second case, Type-2 blocks use sublinear pooling in the model as Fig.1 shows. Results in TABLE I demonstrate that sublinear pooling not only decrease the detection error rate but also decrease the performance gap between training set and testing set, indicating our new pooling method improve the model's generalization ability to detect steganagraphic algorithms.

\begin{figure*}[t]
   \centering
   \begin{subfigure}{.45\textwidth}
     \centering
     \includegraphics[height=6.1cm, width=8.7cm]{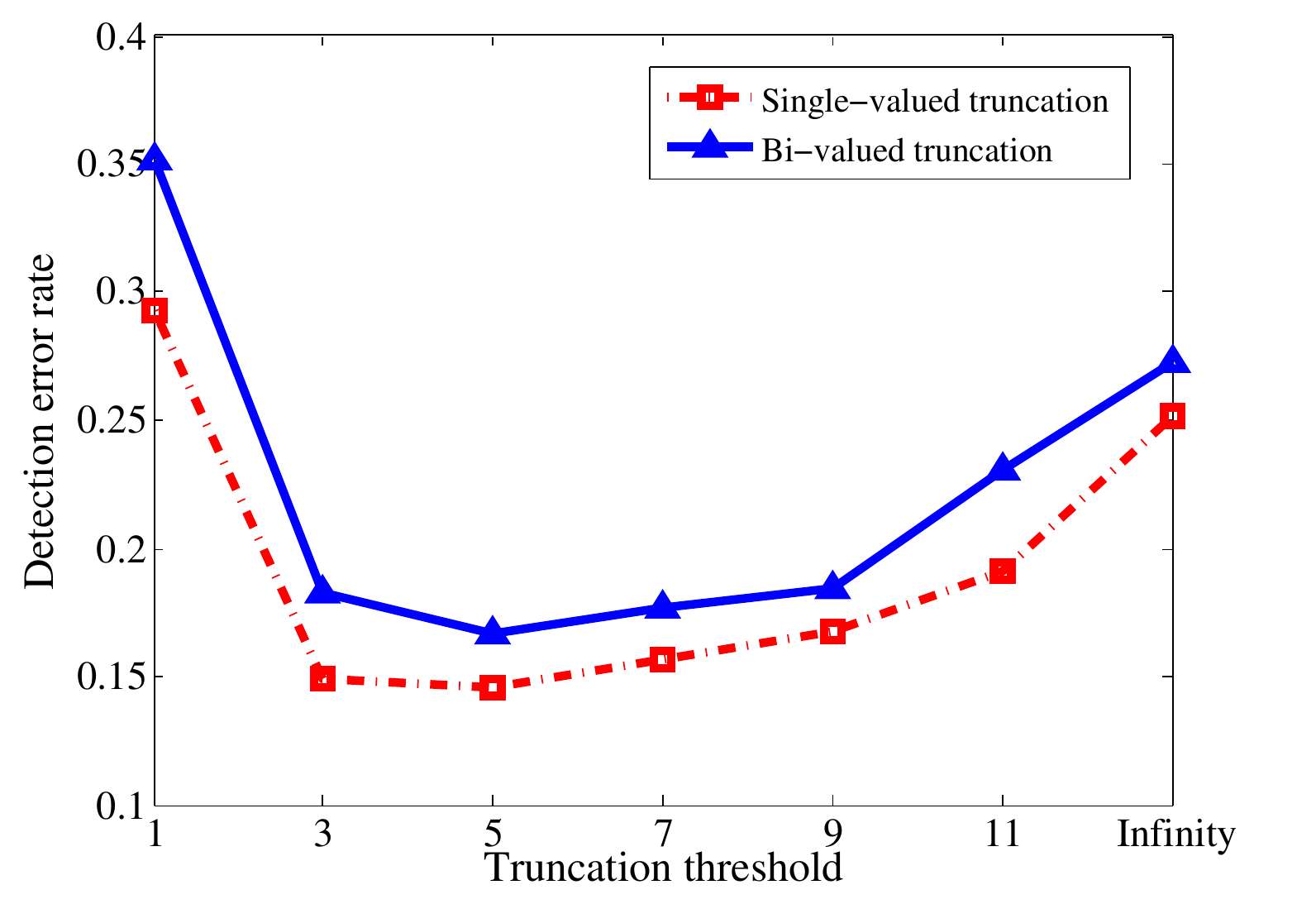}
     \caption{Detection error rates at different truncation values}
   \end{subfigure}
   \begin{subfigure}{.45\textwidth}
     \centering
     \includegraphics[height=6.1cm, width=8.7cm]{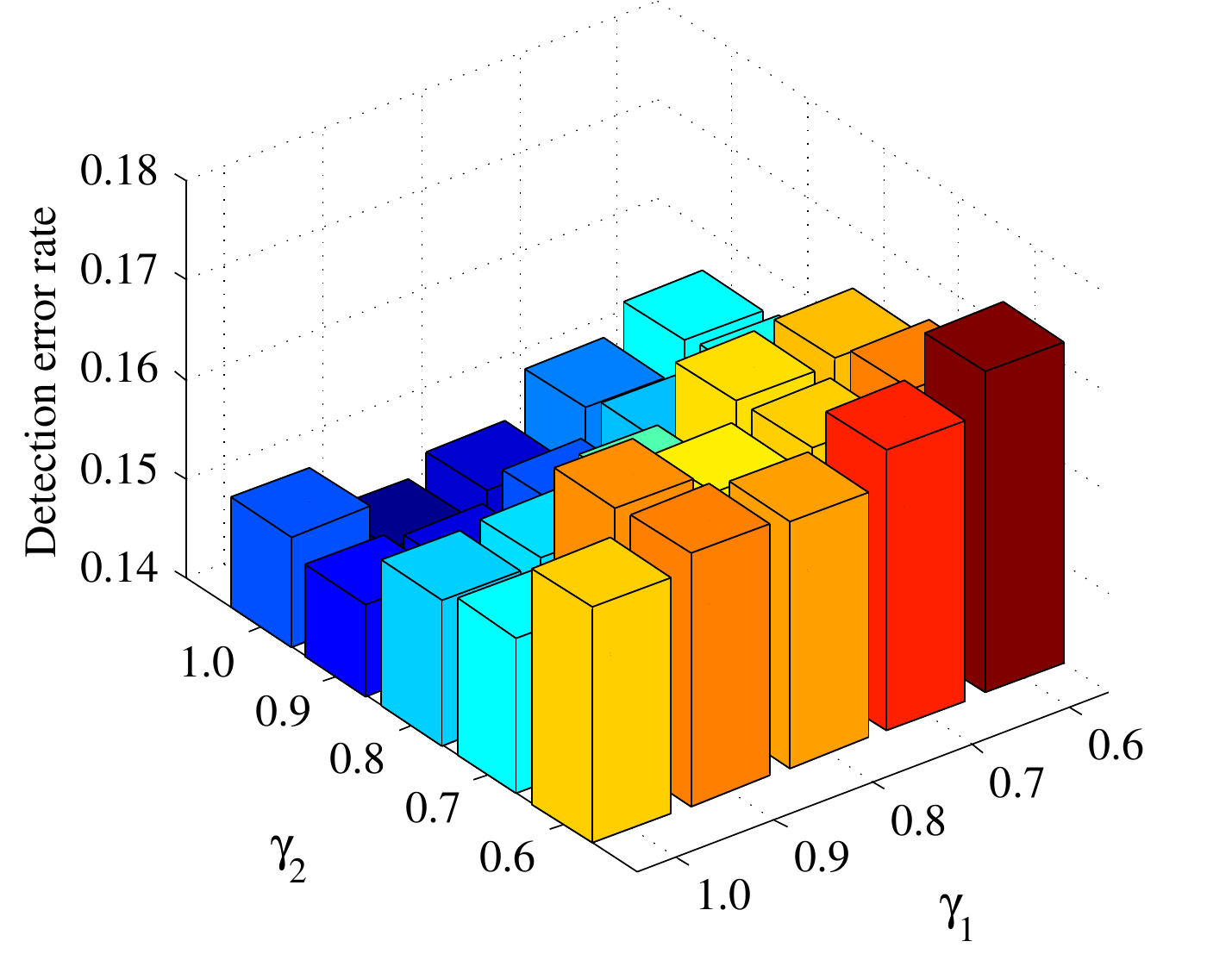}
     \caption{Detection error rates at different configuration of ($\gamma_{1},\gamma_{2}$)}
   \end{subfigure}
   \caption{Ablation study to proposed CIS-Net. (a) Detection error rates of CIS-Net with STL and BTL at different truncation thresholds. (b) Detection error rates of CIS-Net at different configuration of $(\gamma_{1}, \gamma_{2})$. }
\end{figure*}

\section{Experiments}
In this section, we conduct extensive experiments to demonstrate the effectiveness of proposed CIS-network for image steganalysis. At first, we introduce implementation details of the proposed model, including parameter setting, optimization method and model training strategy. Secondly, we validate the proposed model on challenging steganographic algorithms and compare it with the state of the art steganalytic methods. Thirdly, we use Class Activation Mapping (CAM) [38] to demonstrate that the proposed CIS-Net can extract the selection channel information effectively even no such information is provided in the model learning phase. Finally, we conduct experiments to validate the effectiveness of proposed network when the database and  steganographic algorithms used in training mismatch that used in testing.

\subsection{Steganographic algorithms and Database}
We use the BOSSbase 1.01 database [44] which contains 10,000 uncompressed natural images  with the size of $512 \times 512$ in all following experiments. For performance evaluation, the detection error rate $P_{E}$ [11-13] is utilized to measure the detection ability of steganalytic algorithms:
\begin{equation}
  P_{E} = min_{P_{FA}}\frac{1}{2}(P_{MD}+P_{FA})
\end{equation}
where $P_{MD}$ and $P_{FA}$ represents the miss detection probability and the false alert probability respectively. Since image steganalysis is a detection problem, we also evaluate the performance of different steganalytic methods with ROC curves on selected payloads.

Our experiments are conducted on states of the art steganographic schemes. Three representatives of adaptive steganographic algorithms, including the Wavelet Obtained Weights steganography (WOW) [5], S-UNIWARD [6], and the HIghpass Low-pass Low-pass steganography (HILL) [7] are adopted for performance evaluation. For all steganographic algorithms, we use the MATLAB version rather than the C++ implementation to avoid the problem as [45] that all images are embedded with a same key.

\subsection{Implementations}
We implement our model using Pytorch platform and the source code can be found at this link. In the implementation, all the weights in the model except the last fully connected layer are initialized by He's  ``improved Xavie'' [46] method:
\begin{equation}
   W_{ij} \sim \mathcal{N}\left(0,\frac{2}{o_{n}}\right)
\end{equation}
where $\mathcal{N}(\cdot,\cdot)$ denotes the Gaussian distribution, $o_{n}$ represents the number of output channels of the convolutional layer. For the fully connected layer, we utilize zero-mean Gaussian random variable to initialize the weights but the variance is set to 0.01. This setting is to avoid that the fully connected layer has a large variance when the ``improved Xavie'' is used for initialization (the variance is 1.0), which may make the model training unstable and hard to converge. The Adam optimizer [47] is used to update the model's parameters in the learning phase. The mini-batch size is set to 16, which contains 8 cover images and their 8 corresponding stego images.

{\setlength{\abovecaptionskip}{2pt}
 \setlength{\belowcaptionskip}{-2pt}
\begin{table*}[t]
  \centering
  \renewcommand\arraystretch{1.2}
  \caption{Performance comparisons between proposed network and SRM, maxSRM on WOW, S-UNIWARD and HILL steganography at five different payloads. The BOSSbase 1.01 dataset is used for validation. }
  \resizebox{14.5cm}{!} {
  \begin{tabular}{ c  c  c  c  c  c  c  }
  \hline
    \textbf{Steganography} & \textbf{Detection algorithm} & 0.1 bpp & 0.2 bpp & 0.3 bpp & 0.4 bpp & 0.5 bpp \\ \hline
    \multirow{3}{*}{\textbf{WOW}} & SRM + ensemble & 40.26\% & 32.10\% & 25.53\% & 20.60\% & 16.83\%  \\
                         & maxSRMd2 + ensemble & 29.97\% & 23.39\% & 18.86\% & 15.43\% & 13.06\% \\
                         & The proposed network & \textbf{29.08\%} & \textbf{21.03\%} & \textbf{15.96\%} & \textbf{12.13\%} & \textbf{9.30\%} \\ \hline
    \multirow{3}{*}{\textbf{S-UNIWARD}} & SRM + ensemble & 40.24\% & 31.99\% & 25.71\% & 20.37\% & 16.40\%\\
                         & maxSRMd2 + ensemble & 36.60\% & 28.86\% & 23.60\% & 19.08\% & 15.51\% \\
                         & The proposed network & \textbf{35.28\%} & \textbf{26.21\%} & \textbf{19.64\%} & \textbf{14.62\%} & \textbf{10.73\%} \\ \hline
    \multirow{3}{*}{\textbf{HILL}} & SRM + ensemble & 43.64\% & 36.11\% & 29.96\% & 24.82\% & 20.55\% \\
                         & maxSRMd2 + ensemble & 37.71\% & 30.91\% & 25.73\% & 21.84\% & 18.14\% \\
                         & The proposed network & \textbf{36.82\%}  & \textbf{28.83\%} & \textbf{22.67\%} & \textbf{18.10\%} & \textbf{14.78\%} \\ \hline
  \end{tabular}
  }
\end{table*}}

\begin{figure*}[t]
   \centering
   \begin{subfigure}{.32\textwidth}
     \centering
     \includegraphics[height=4.5cm, width=5.7cm]{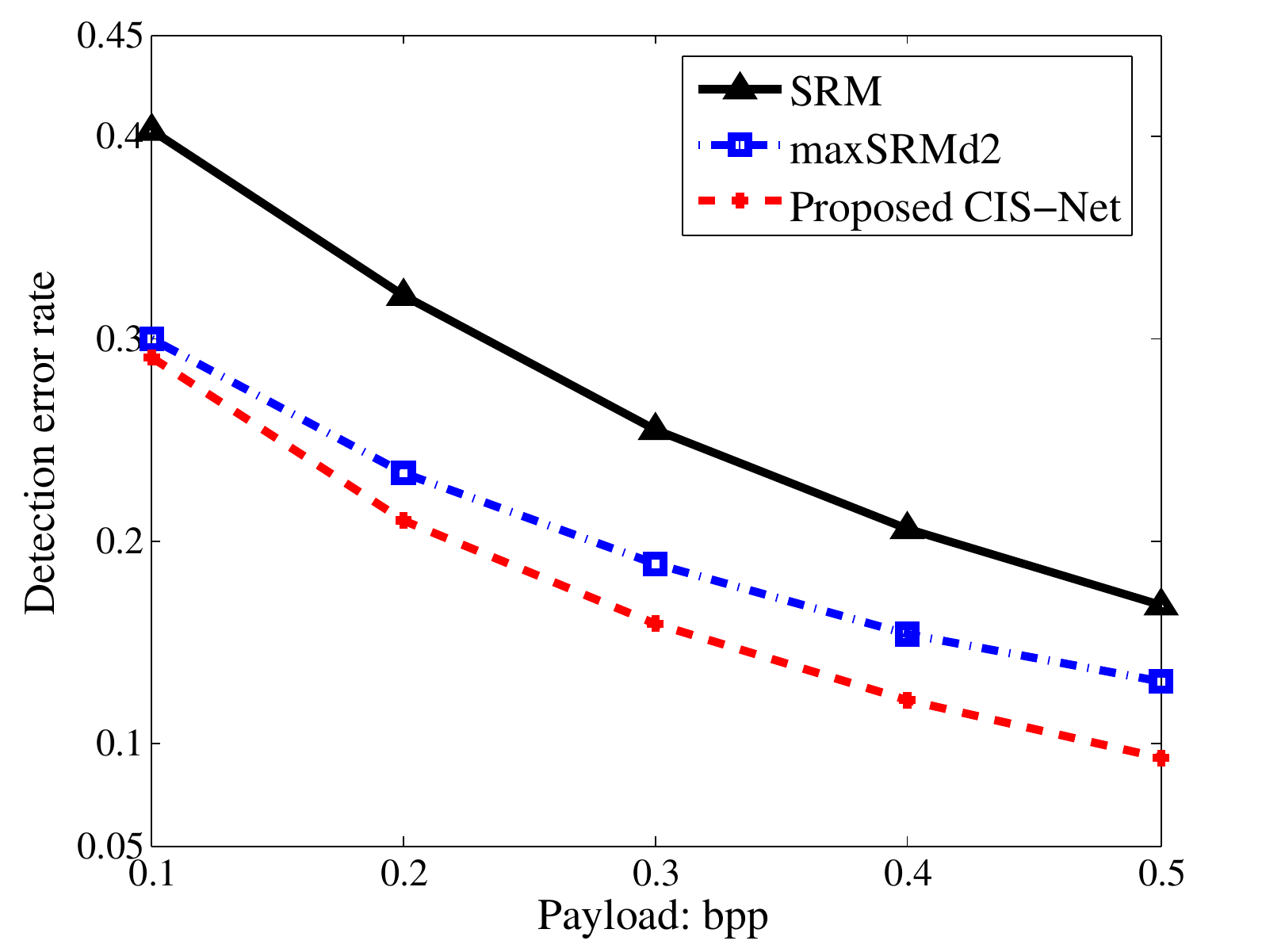}
     \caption{WOW steganography}
    \end{subfigure}
    \begin{subfigure}{.32\textwidth}
     \centering
     \includegraphics[height=4.5cm, width=5.7cm]{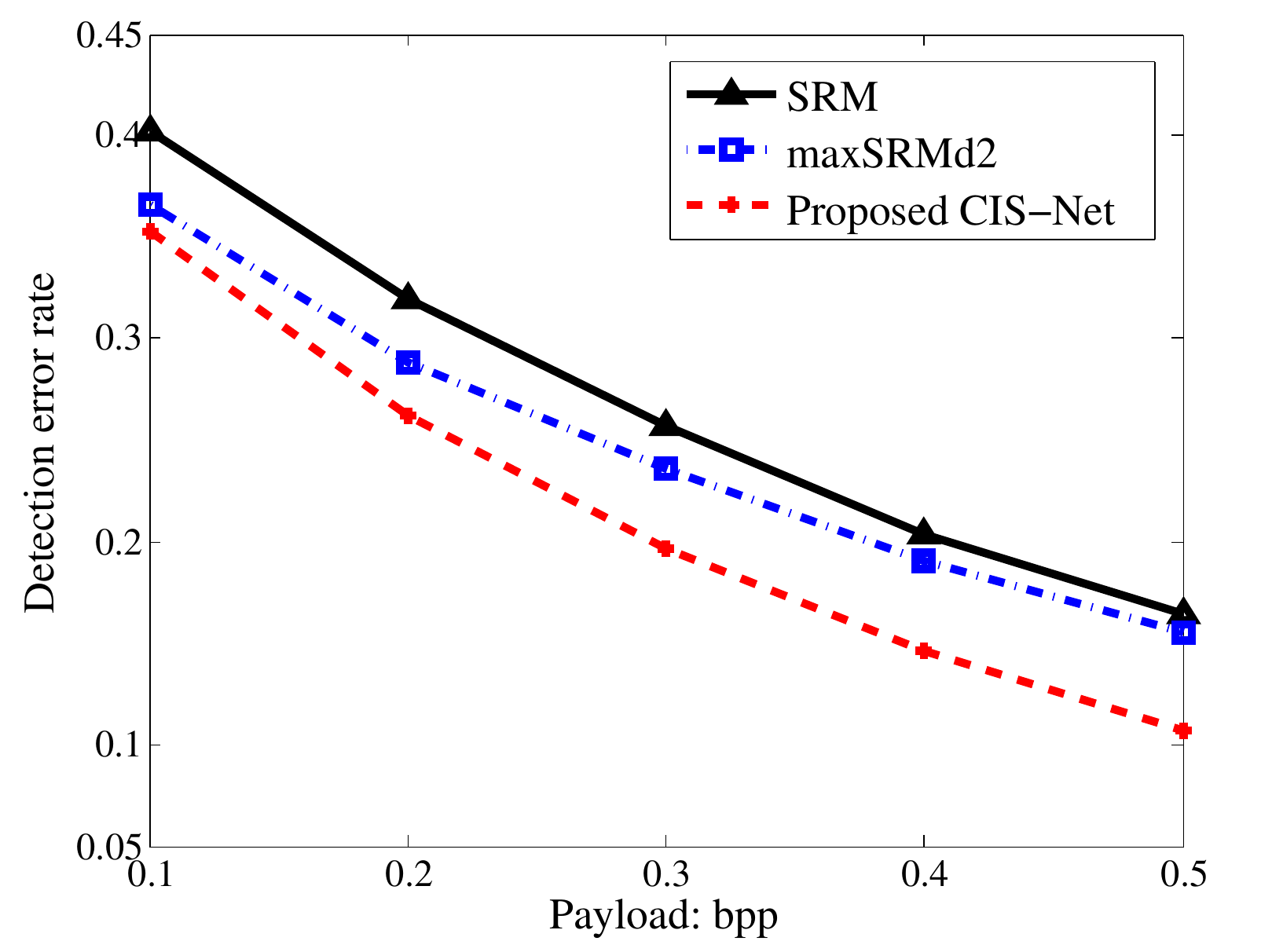}
     \caption{S-UNIWARD steganography}
    \end{subfigure}
    \begin{subfigure}{.32\textwidth}
     \centering
     \includegraphics[height=4.5cm, width=5.7cm]{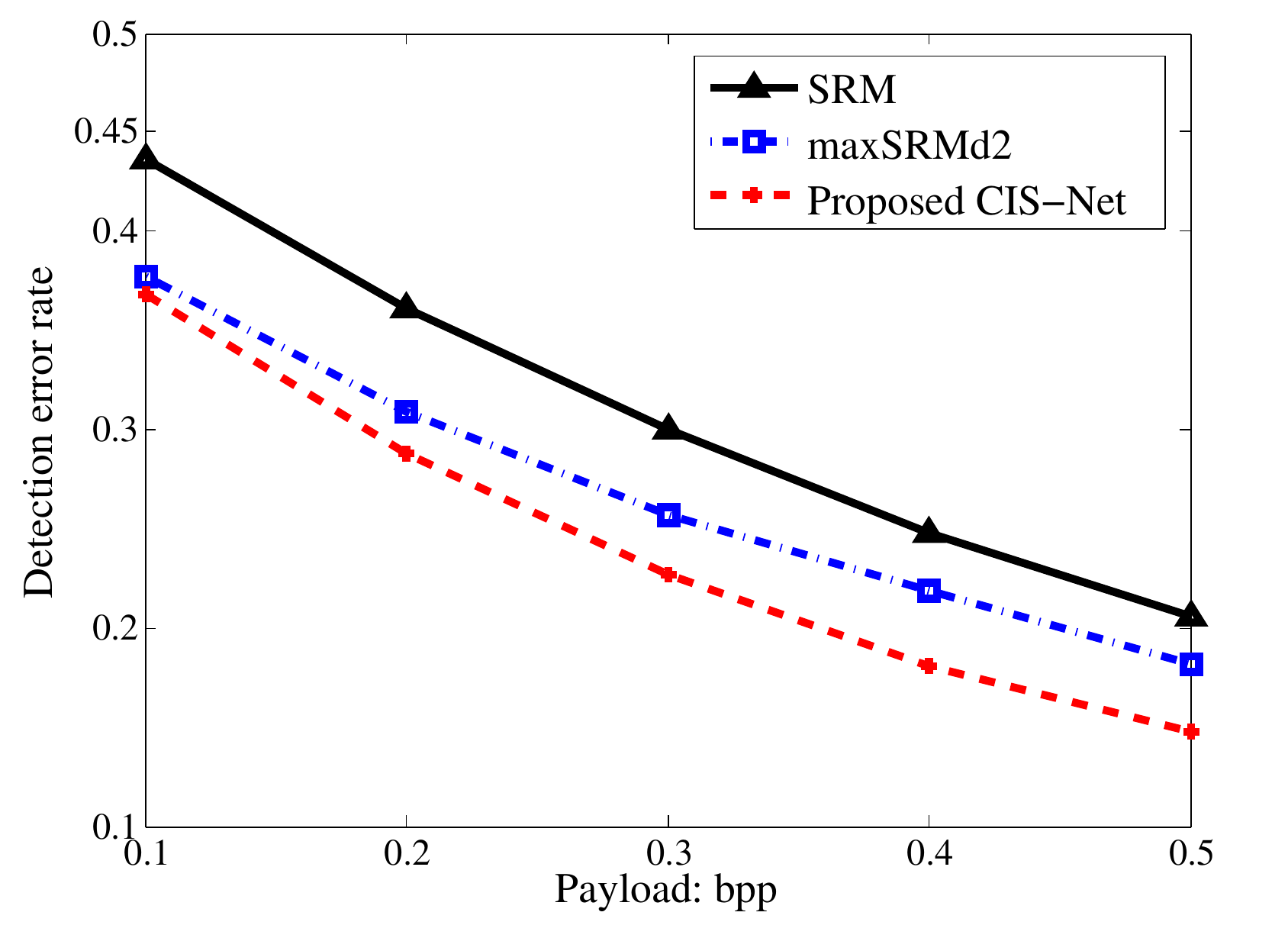}
     \caption{HILL steganography}
    \end{subfigure}
   \caption{Detection error rates for SRM, maxSRM and the proposed network for three different steganographic algorithms at five different payloads. }
\end{figure*}

\begin{figure*}[t]
   \centering
   \begin{subfigure}{.32\textwidth}
     \centering
     \includegraphics[height=4.5cm, width=5.7cm]{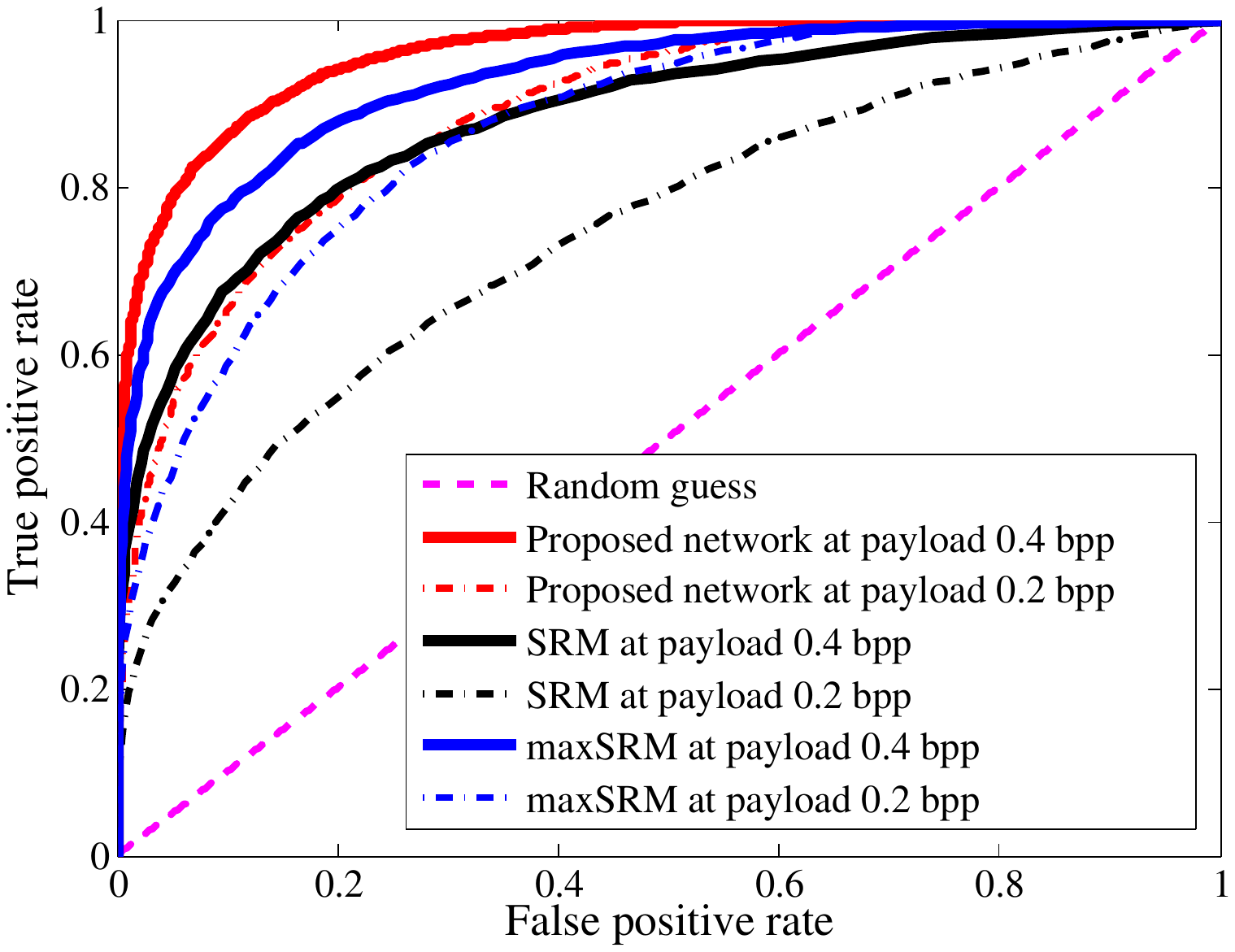}
     \caption{WOW steganography}
   \end{subfigure}
   \begin{subfigure}{.32\textwidth}
     \centering
     \includegraphics[height=4.5cm, width=5.7cm]{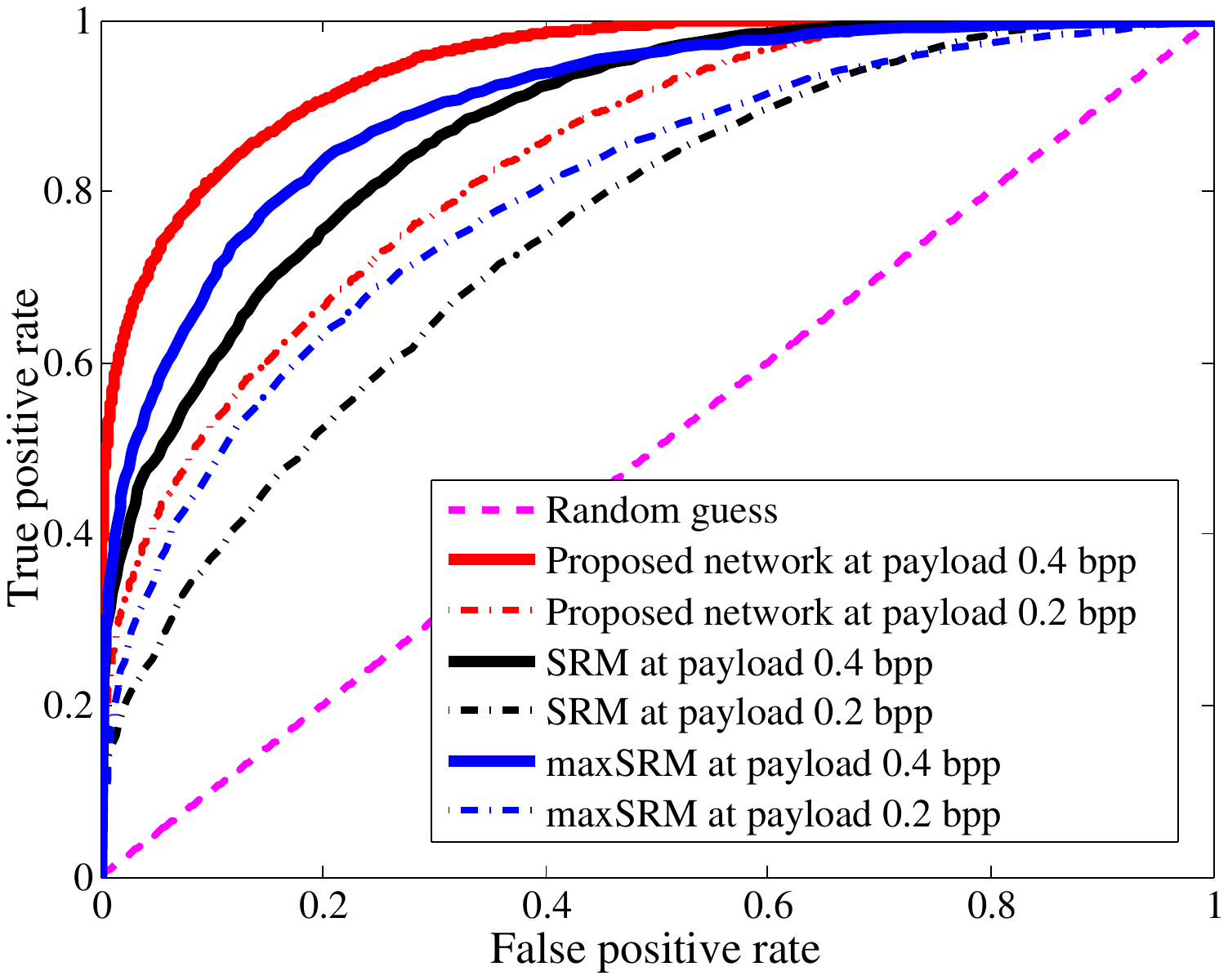}
     \caption{S-UNIWARD steganography}
   \end{subfigure}
   \begin{subfigure}{.32\textwidth}
     \centering
     \includegraphics[height=4.5cm, width=5.7cm]{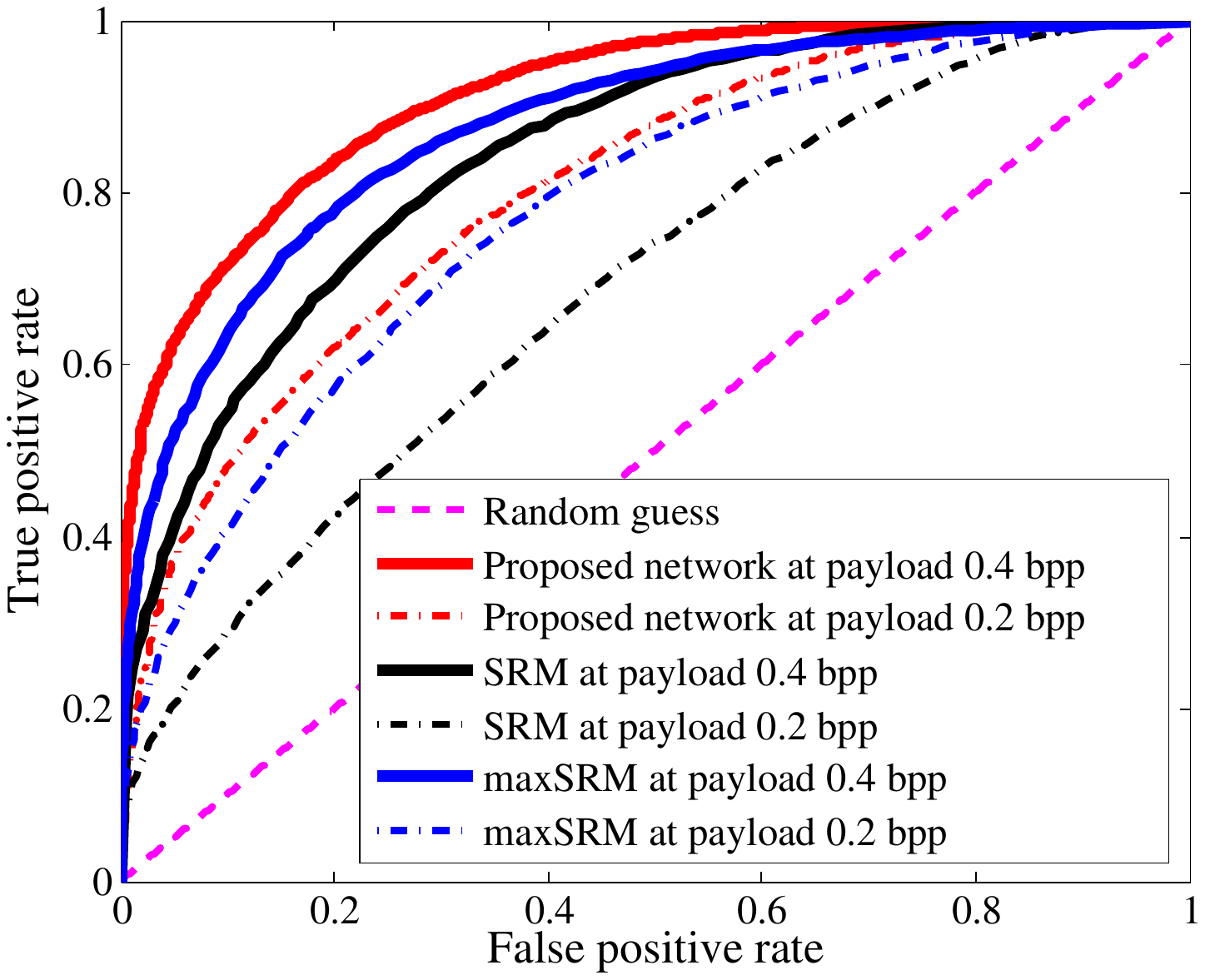}
     \caption{HILL steganography}
   \end{subfigure}
   \caption{ROC curves of SRM, maxSRM and the proposed network for three different steganographic algorithms at payloads 0.4 bpp and 0.2 bpp. }
\end{figure*}

In our model, all convolutional layers including fully connected layer contain bias. Unlike general methods that just initialize biases with random values, we calculate the bias of each convolutional layer based on input cover-stego pairs in the initialization stage. For the $n$-th convolutional layer, its bias $b_{n}$ is set by the following formula:
\begin{equation}
   b_{n} = -\frac{1}{|\mathcal{S}|} \sum_{i \in \mathcal{S}} E \left[ vec\left( \bm{c}_{n}^{xi} \right) + vec\left( \bm{c}_{n}^{yi} \right) \right]
\end{equation}
where $\bm{c}_{n}^{xi}$ and $\bm{c}_{n}^{yi}$ denote the $i$-th feature map of the $n$-th convolutional layer for the cover image $x$ and stego image $y$. $vec(\cdot)$ represents the vectorization operator and $E(\cdot)$ is the expectation. $\mathcal{S}$ is the set of cover/stego images for initializing the bias of convolutional/fully-connected layers. Actually, this is a mean-only version of shared normalization proposed in [25]. The advantage of such is that it can conceal the non-zero mean introduced by the STL and make all feature map elements distributed across zero, leading to fast convergence. In our experiment, the size of $\mathcal{S}$ is set to 100, which contains 50 randomly selected cover images and their corresponding stegos from training set.

{\setlength{\abovecaptionskip}{2pt}
 \setlength{\belowcaptionskip}{-2pt}
\begin{table*}[t]
  \centering
  \renewcommand\arraystretch{1.2}
  \caption{Performance comparisons between proposed network and several state of the arts CNN models on S-UNIWARD and HILL at five different payloads. The BOSSbase 1.01 dataset is used for validation. }
  \resizebox{14.5cm}{!} {
  \begin{tabular}{ c  c  c  c  c  c  c  }
  \hline
    \textbf{Steganography} & \textbf{CNN model} & 0.1 bpp & 0.2 bpp & 0.3 bpp & 0.4 bpp & 0.5 bpp \\ \hline
    \multirow{5}{*}{\textbf{S-UNIWARD}} & Xu-network & 40.57\% & 33.33\% & 26.32\% & 19.88\% & 16.46\%\\
                         & Ye-network & 40.29\% & 33.51\% & 25.62\% & 22.64\% & 17.64\% \\
                         & SN-network & \textbf{35.21\%} & 26.82\% & 20.71\% & 16.53\% & 12.71\% \\
                         & ReST-Net (SRM) & 35.85\% & 31.27\% & 23.56\% & 15.72\% & 13.83\% \\
                         & The proposed network & 35.28\% & \textbf{26.21\%} & \textbf{19.64\%} & \textbf{14.62\%} & \textbf{10.73\%} \\ \hline
    \multirow{5}{*}{\textbf{HILL}} & Xu-network & 41.07\% & 33.25\% & 26.86\% & 21.31\% & 18.18\% \\
                         & Ye-network  & 43.55\% & 34.65\% & 27.98\% & 23.08\% & 21.14\% \\
                         & SN-network & 36.86\% & 29.63\% & 23.60\% & 19.87\% & 16.29\% \\
                         & ReST-Net (SRM) & 38.77\% & 30.87\% & 24.84\% & 19.75\% & 16.53\% \\
                         & The proposed network & \textbf{36.82\%}  & \textbf{28.83\%} & \textbf{22.67\%} & \textbf{18.10\%} & \textbf{14.78\%} \\ \hline
  \end{tabular}
  }
\end{table*}}

\begin{figure*}[t]
   \centering
   \begin{subfigure}{.19\textwidth}
     \centering
     \includegraphics[height=3.4cm, width=3.4cm]{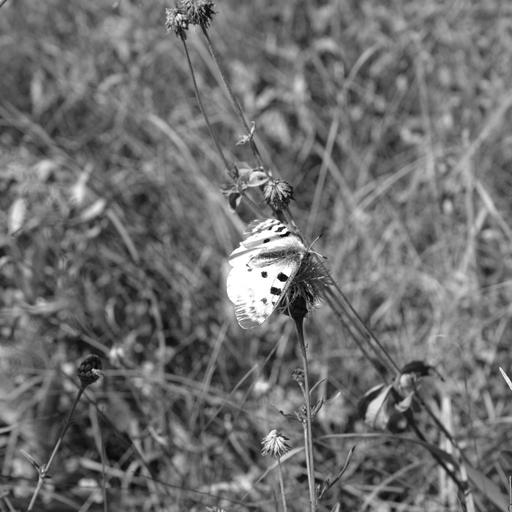}
   \end{subfigure}
   \begin{subfigure}{.19\textwidth}
      \centering
      \includegraphics[height=3.4cm, width=3.4cm]{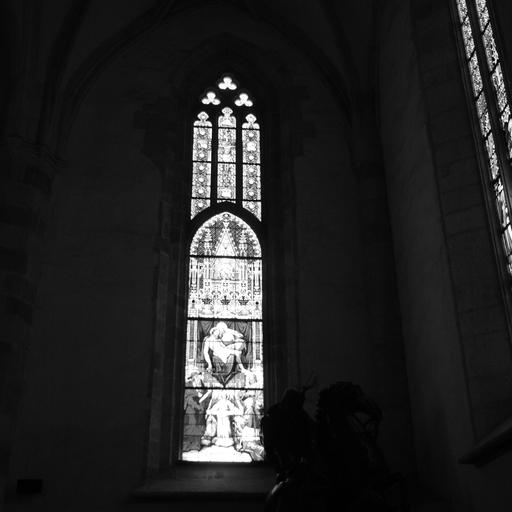}
   \end{subfigure}
   \begin{subfigure}{.19\textwidth}
     \centering
     \includegraphics[height=3.4cm, width=3.4cm]{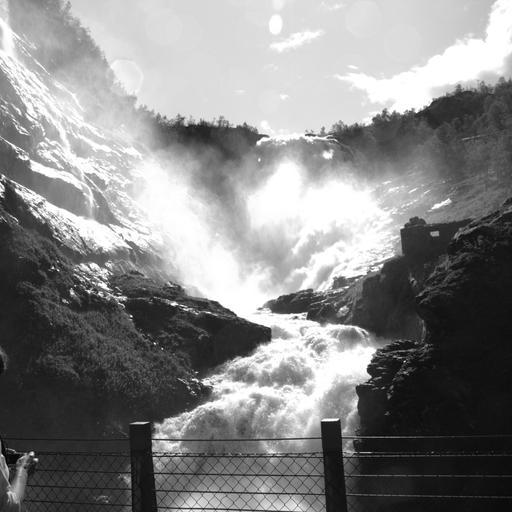}
   \end{subfigure}
   \begin{subfigure}{.19\textwidth}
     \centering
     \includegraphics[height=3.4cm, width=3.4cm]{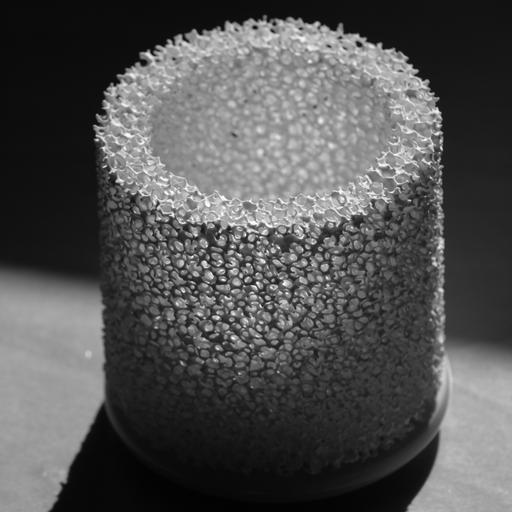}
   \end{subfigure}
   \begin{subfigure}{.19\textwidth}
     \centering
     \includegraphics[height=3.4cm, width=3.4cm]{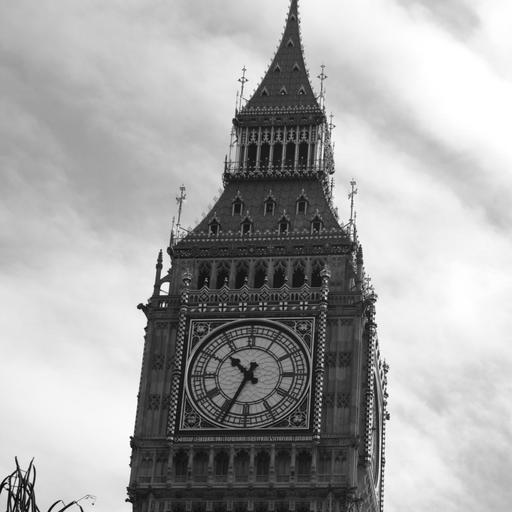}
   \end{subfigure}
   \caption*{(a) Cover images selected from BOSSbase 1.01}

   \begin{subfigure}{.19\textwidth}
     \centering
     \includegraphics[height=3.4cm, width=3.4cm]{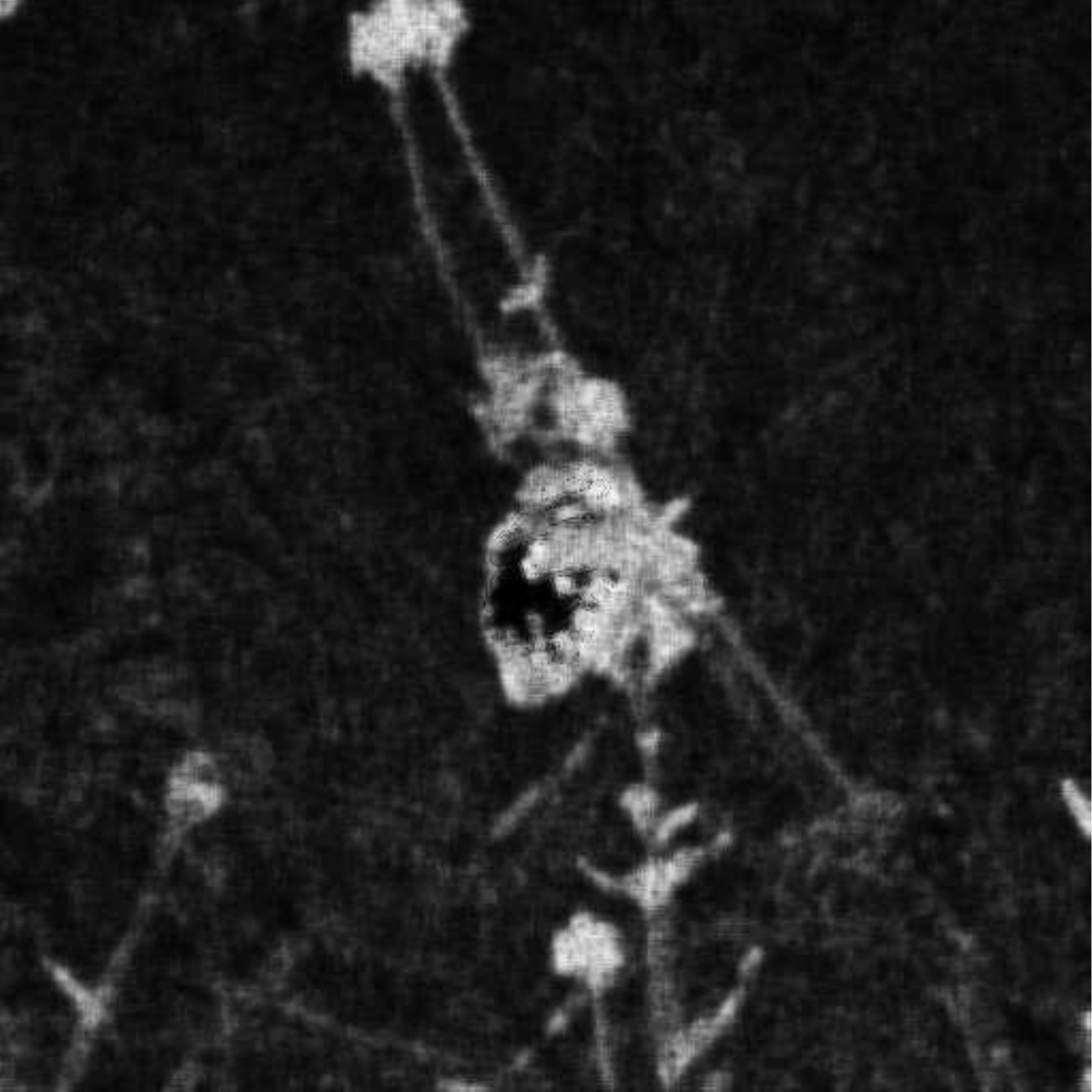}
   \end{subfigure}
   \begin{subfigure}{.19\textwidth}
      \centering
      \includegraphics[height=3.4cm, width=3.4cm]{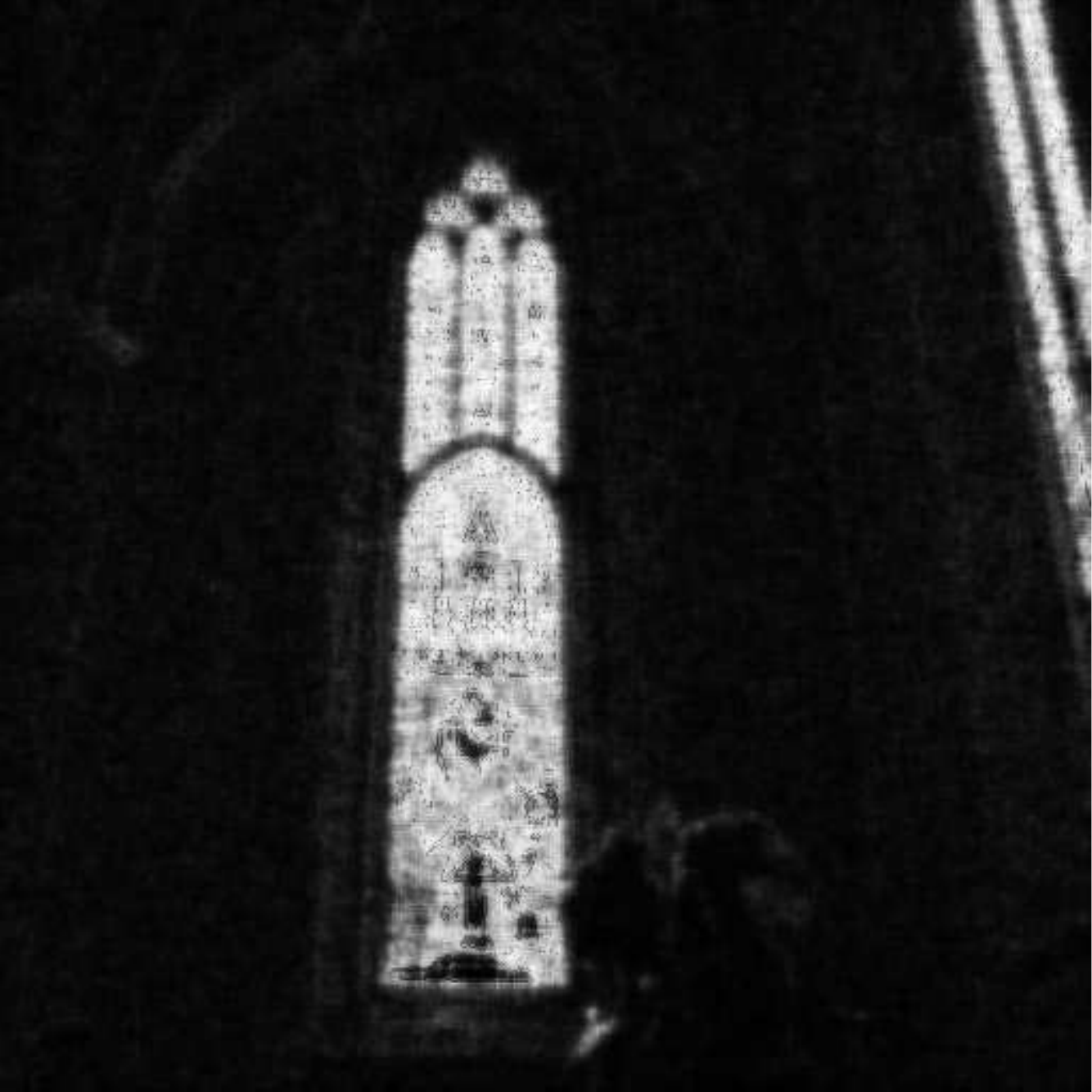}
   \end{subfigure}
   \begin{subfigure}{.19\textwidth}
     \centering
     \includegraphics[height=3.4cm, width=3.4cm]{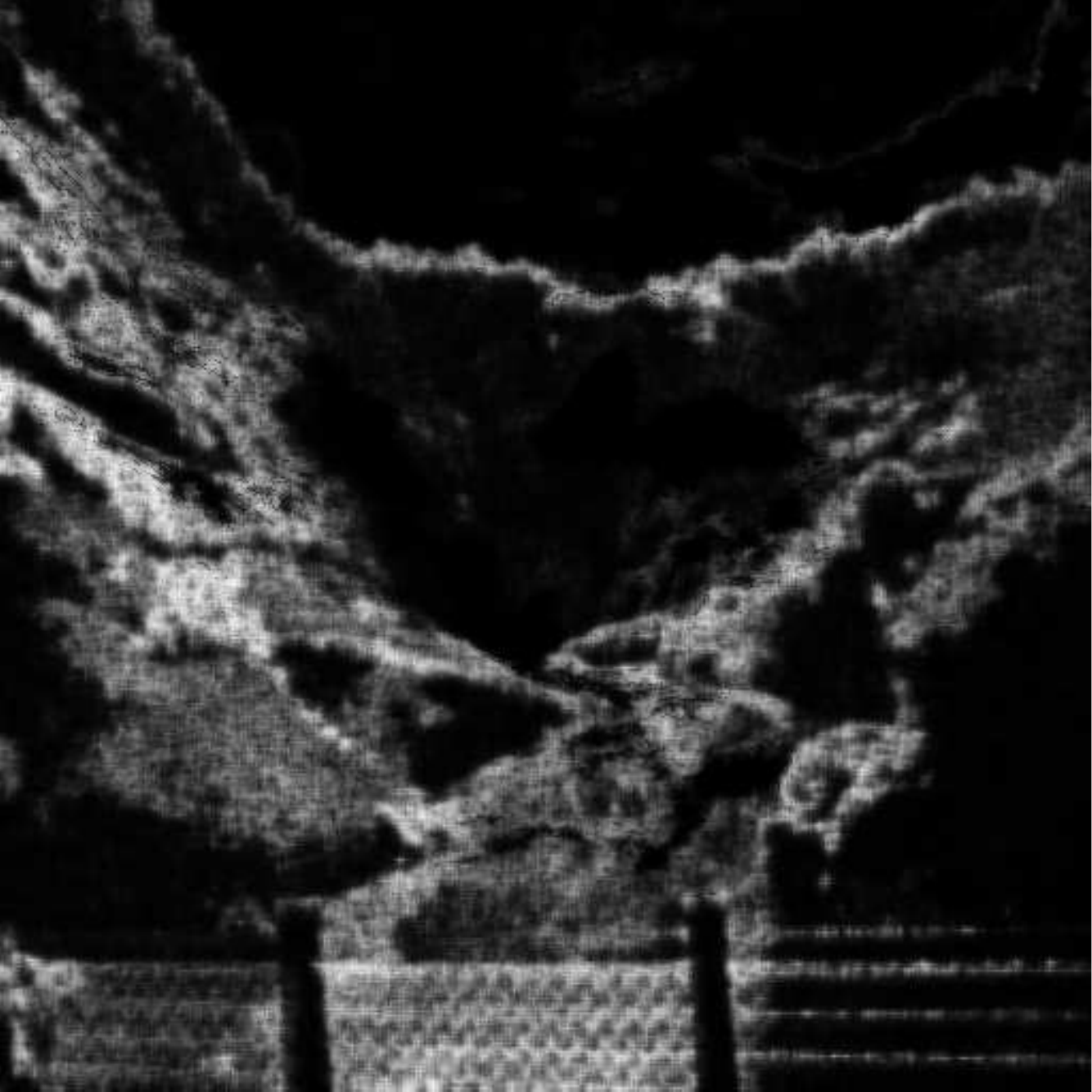}
   \end{subfigure}
   \begin{subfigure}{.19\textwidth}
     \centering
     \includegraphics[height=3.4cm, width=3.4cm]{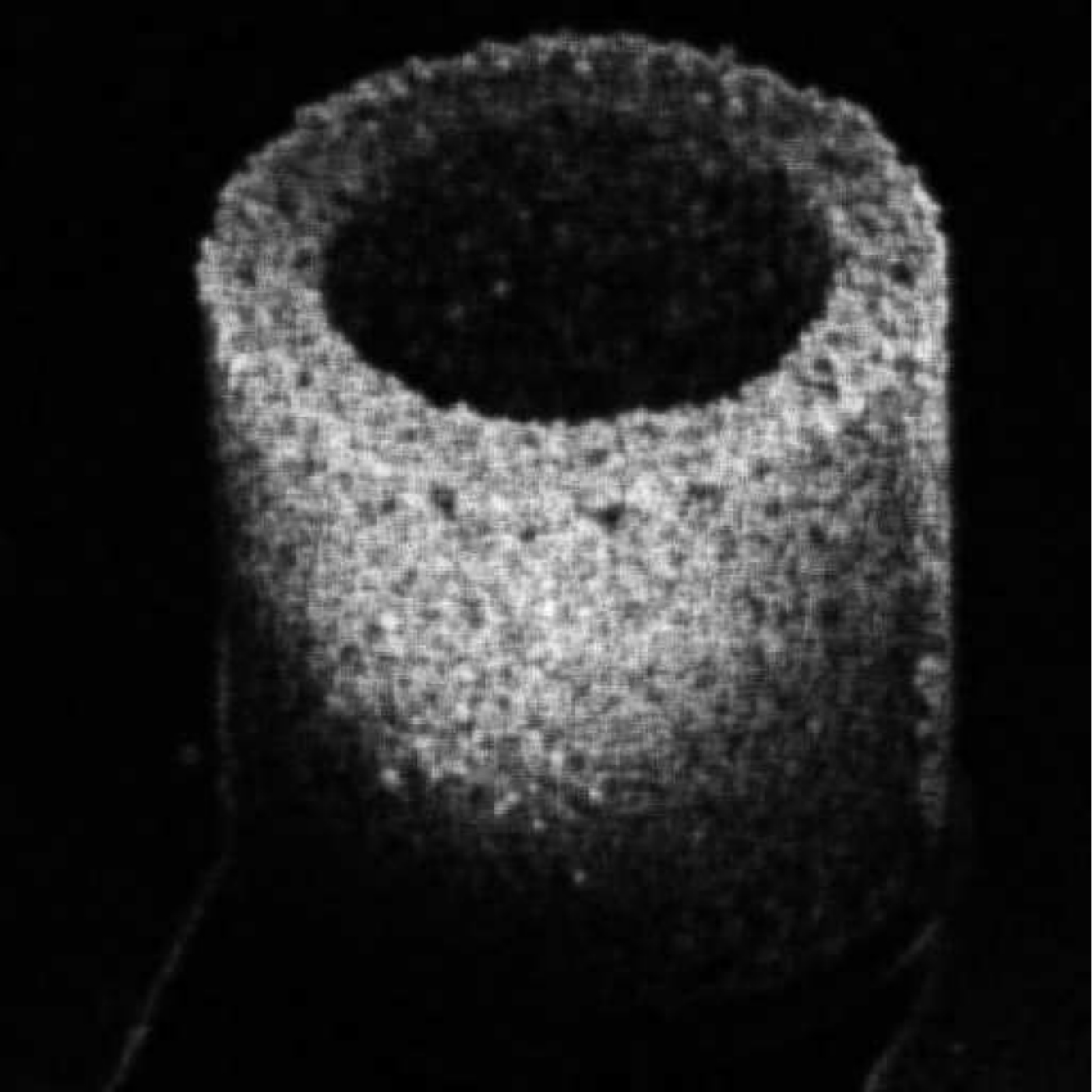}
   \end{subfigure}
   \begin{subfigure}{.19\textwidth}
     \centering
     \includegraphics[height=3.4cm, width=3.4cm]{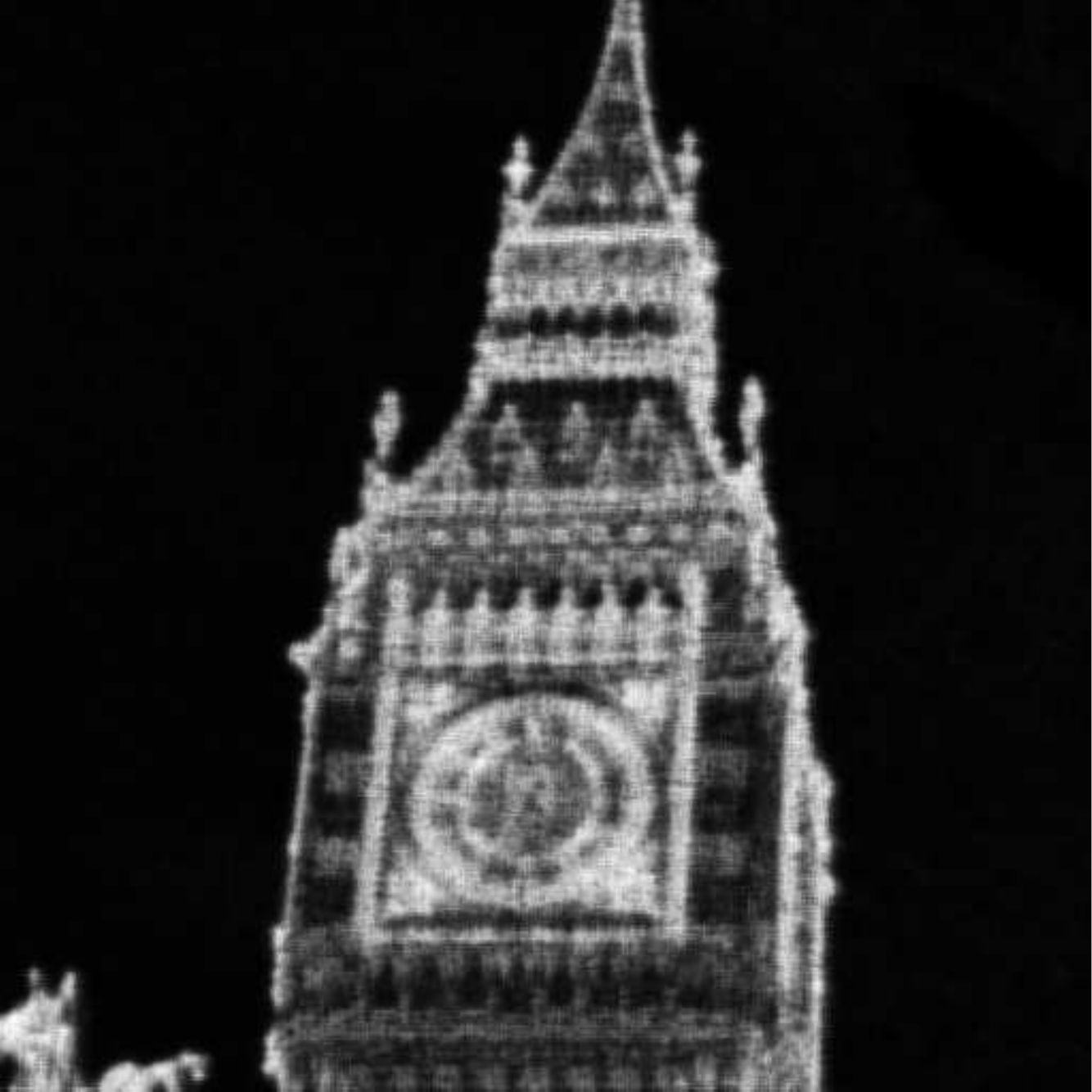}
   \end{subfigure}
   \caption*{(b) Ground truth of embedding probability map}

   \begin{subfigure}{.19\textwidth}
     \centering
     \includegraphics[height=3.4cm, width=3.4cm]{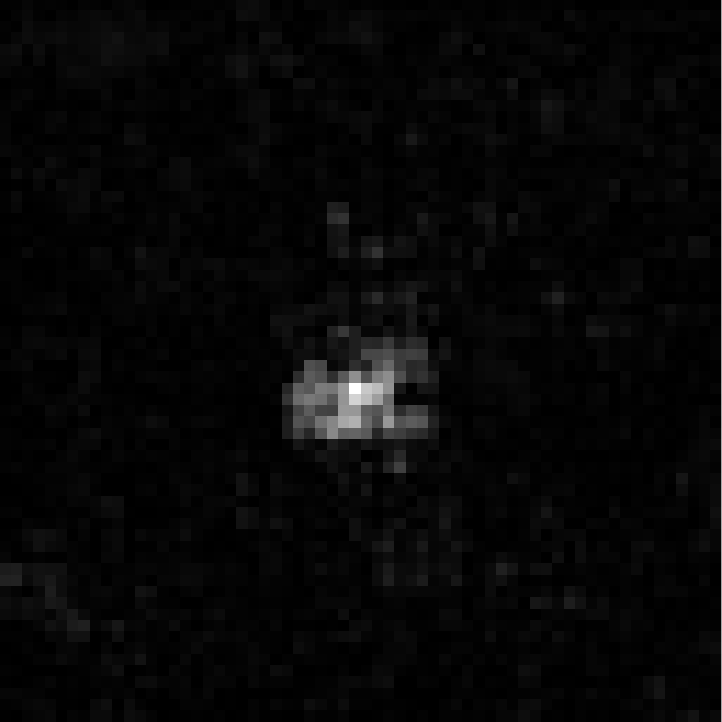}
   \end{subfigure}
   \begin{subfigure}{.19\textwidth}
      \centering
      \includegraphics[height=3.4cm, width=3.4cm]{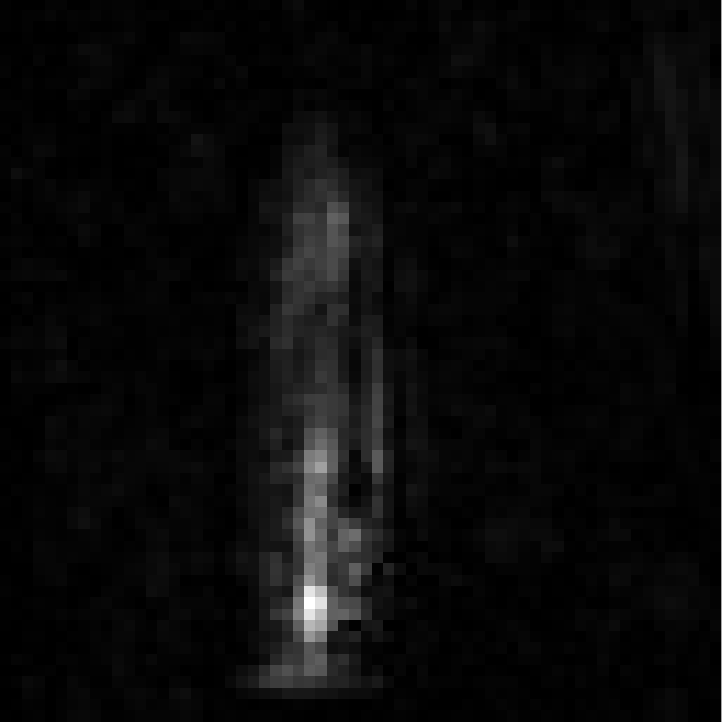}
   \end{subfigure}
   \begin{subfigure}{.19\textwidth}
     \centering
     \includegraphics[height=3.4cm, width=3.4cm]{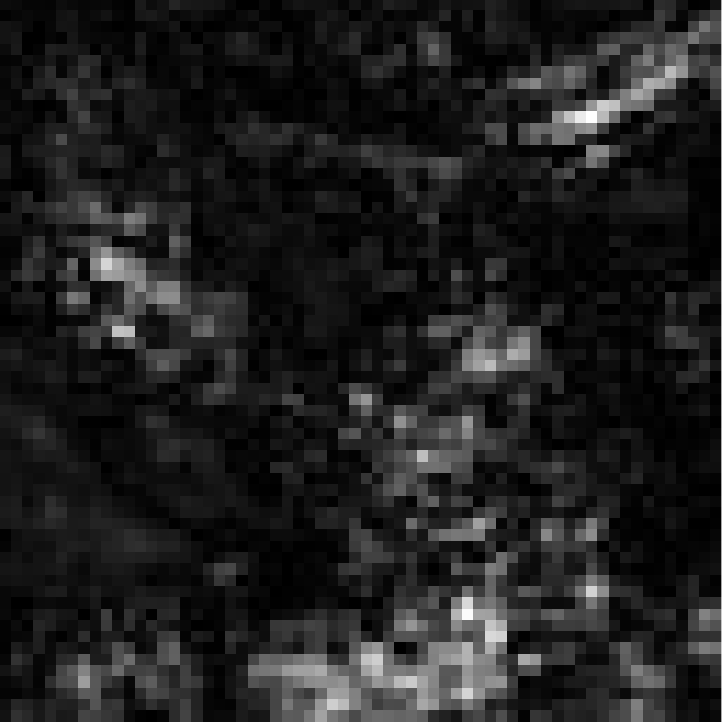}
   \end{subfigure}
   \begin{subfigure}{.19\textwidth}
     \centering
     \includegraphics[height=3.4cm, width=3.4cm]{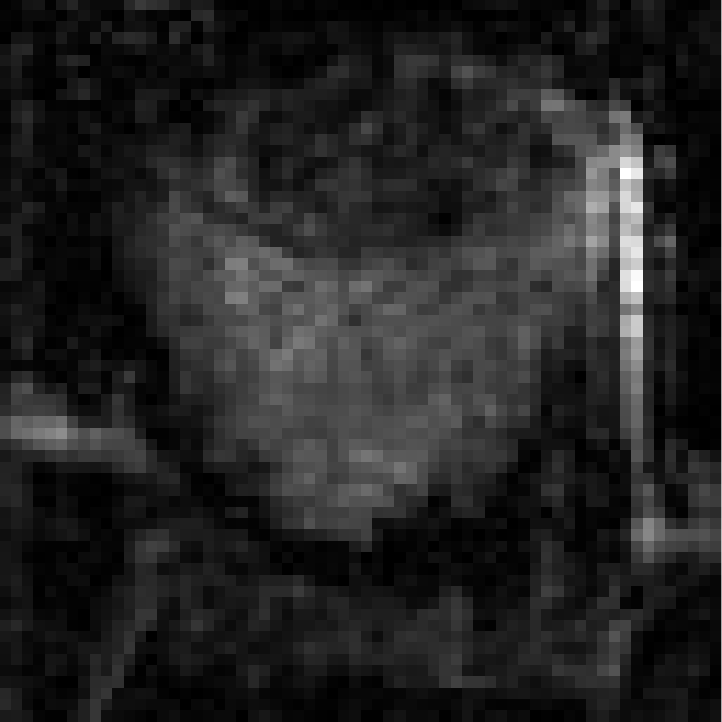}
   \end{subfigure}
   \begin{subfigure}{.19\textwidth}
     \centering
     \includegraphics[height=3.4cm, width=3.4cm]{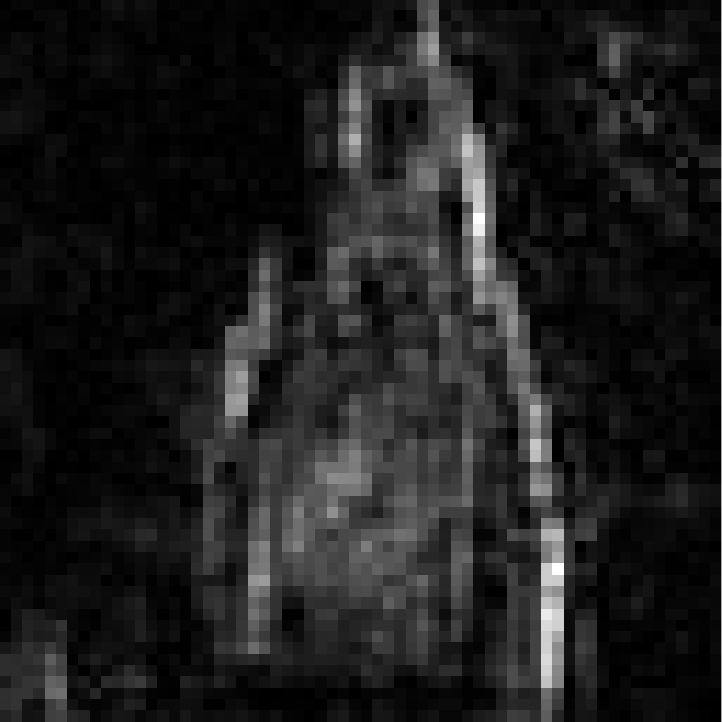}
   \end{subfigure}
   \caption*{(c) Attentional maps extracted from proposed CIS-Net}

   \caption{Comparisons of attentional maps and probability embedding maps on different images. (a) Five selected cover images from BOSSbase 1.01. (b) Ground truth probability embedding maps of selected cover images on on S-UNIWARD steganography at 0.4 bpp. (c) Attentional maps of selected cover images extracted from CIS-Net based on CAM.  }
\end{figure*}

\begin{figure*}[t]
   \begin{subfigure}{.33\textwidth}
     \centering
     \includegraphics[height=5.8cm, width=5.8cm]{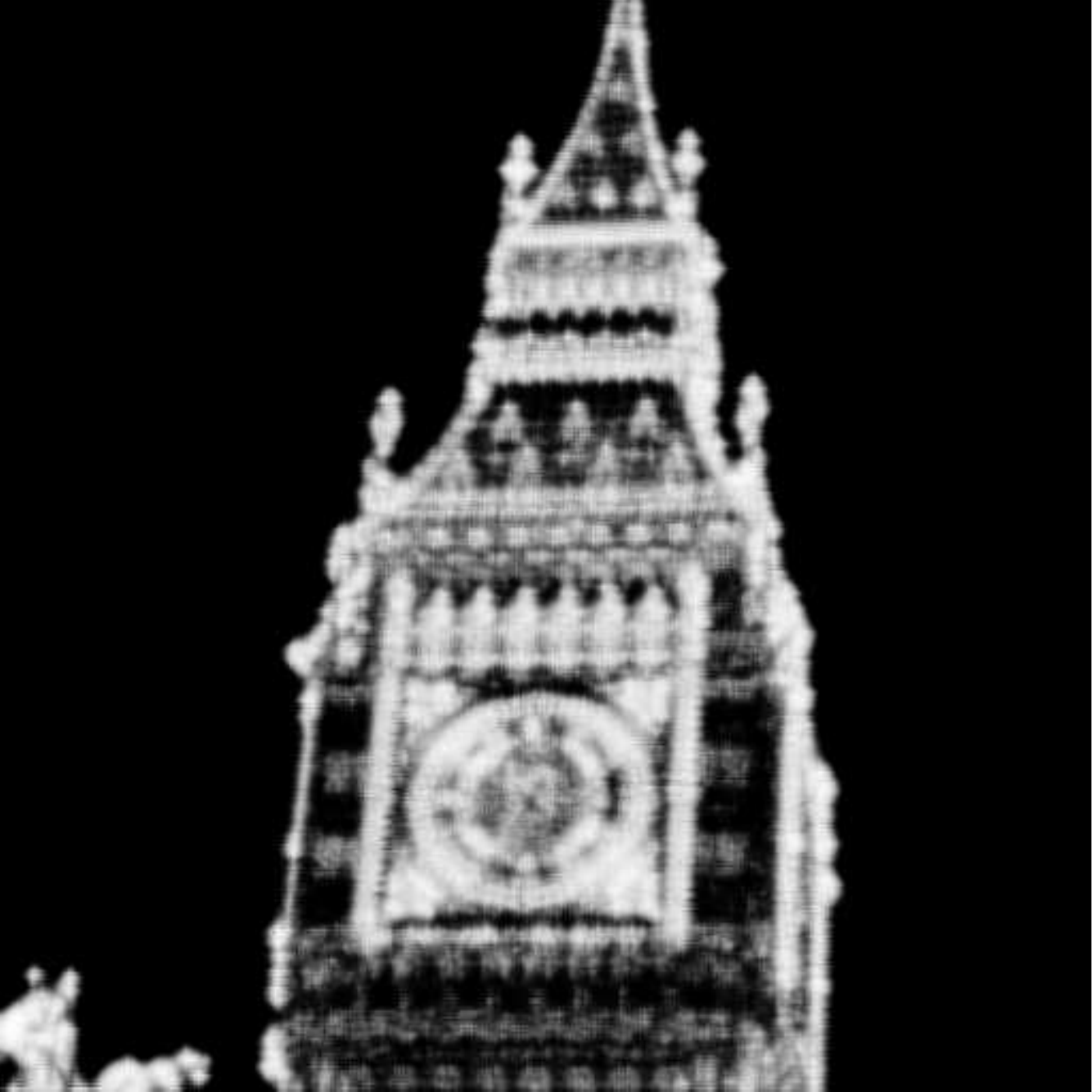}
     \caption{WOW embedding map at 0.4 bpp}
   \end{subfigure}
   \begin{subfigure}{.33\textwidth}
     \centering
     \includegraphics[height=5.8cm, width=5.8cm]{2387_map.pdf}
     \caption{S-UNIWARD embedding map at 0.4 bpp}
   \end{subfigure}
   \begin{subfigure}{.33\textwidth}
     \centering
     \includegraphics[height=5.8cm, width=5.8cm]{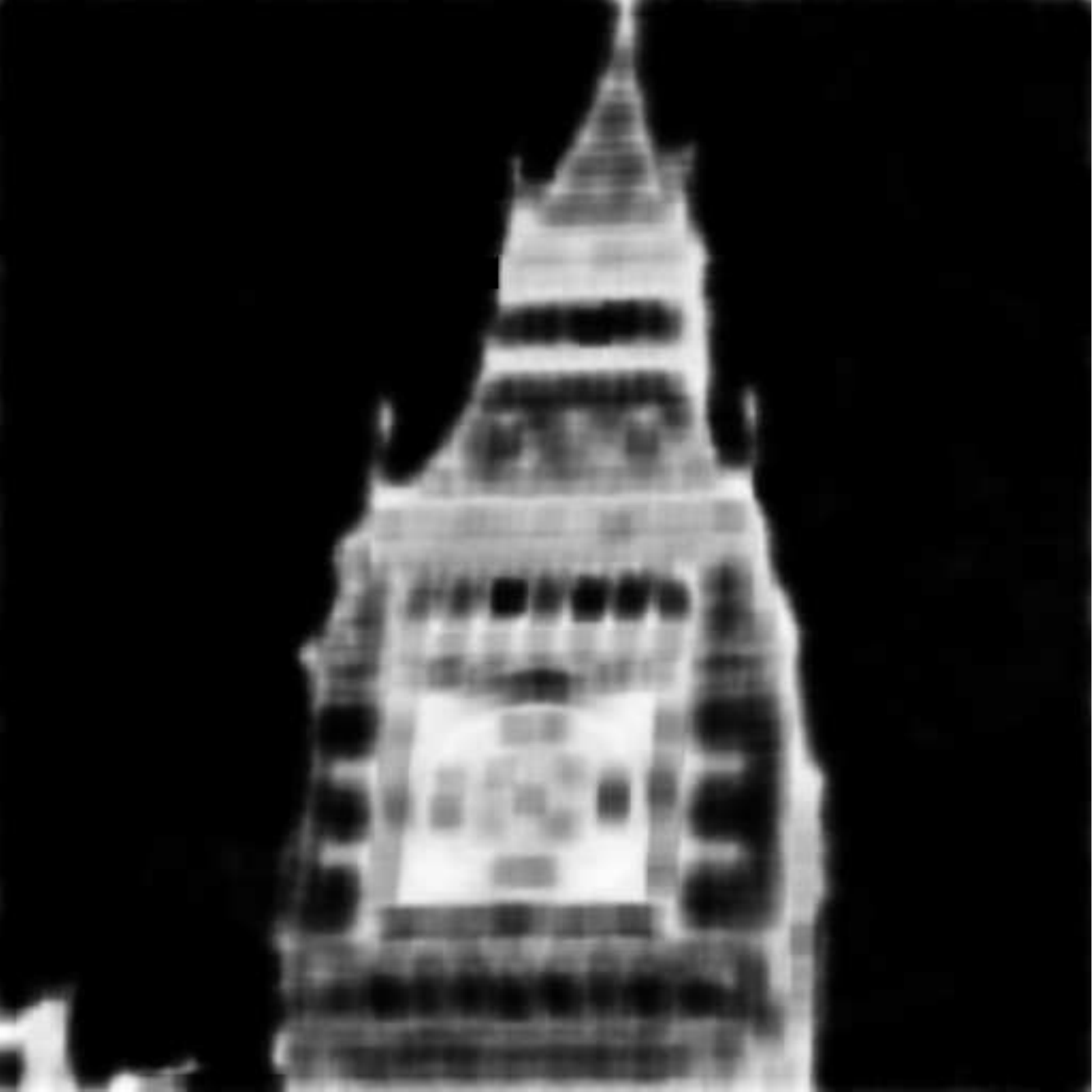}
     \caption{HILL embedding map at 0.4 bpp}
   \end{subfigure}

   \begin{subfigure}{.33\textwidth}
     \centering
     \includegraphics[height=5.8cm, width=5.8cm]{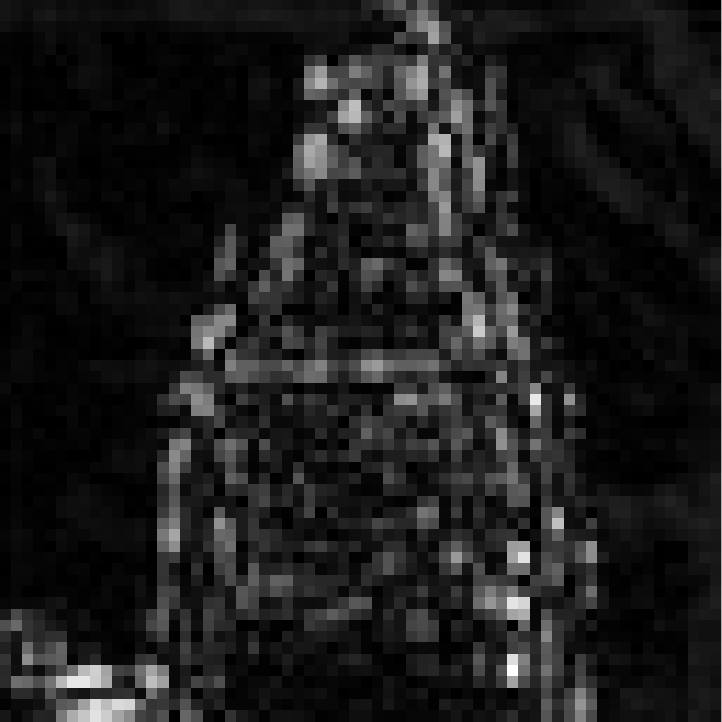}
     \caption{Attentional map of WOW at 0.4 bpp}
   \end{subfigure}
   \begin{subfigure}{.33\textwidth}
     \centering
     \includegraphics[height=5.8cm, width=5.8cm]{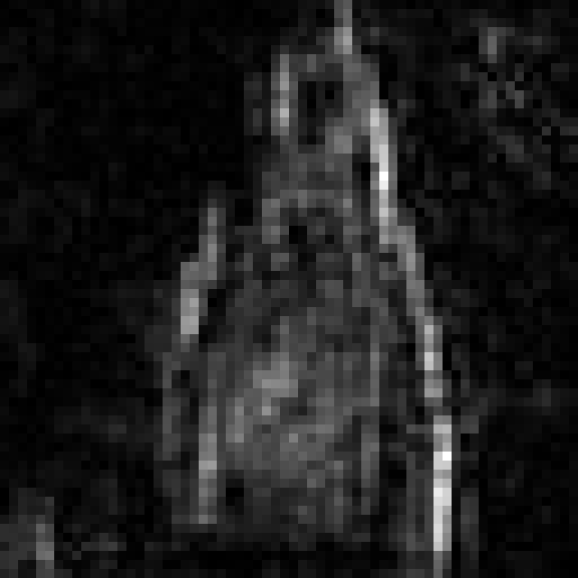}
     \caption{Attentional map of S-UNIWARD at 0.4 bpp}
   \end{subfigure}
   \begin{subfigure}{.33\textwidth}
     \centering
     \includegraphics[height=5.8cm, width=5.8cm]{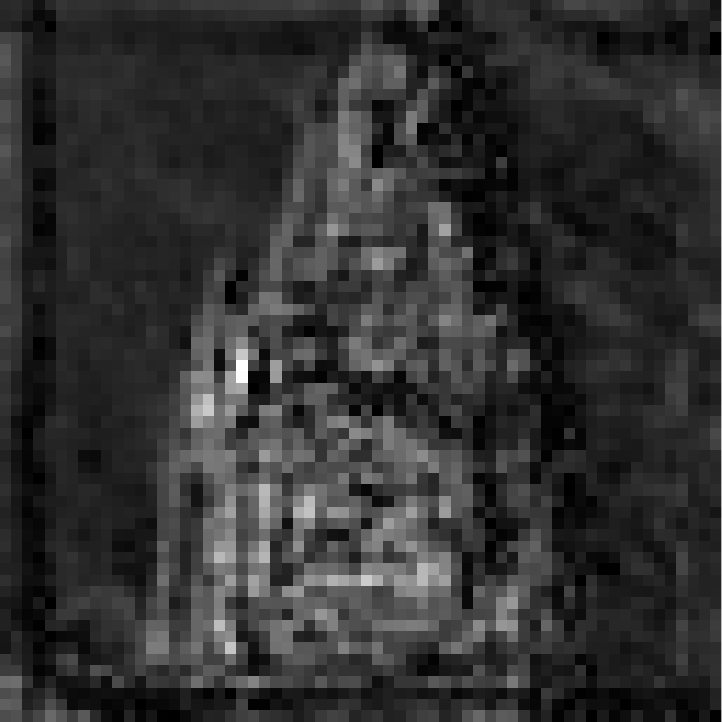}
     \caption{Attentional map of HILL at 0.4 bpp}
   \end{subfigure}
   \caption{Comparisons of probability embedding maps and attentional maps for different steganographic embedding methods. (a) and (d) are probability embedding maps and attentional maps of WOW steganography at 0.4 bpp; (b) and (e) are probability embedding maps and attentional maps of S-UNIWARD steganography at 0.4 bpp; (c) and (f) are probability embedding maps and attentional maps of HILL steganography at 0.4 bpp. }
\end{figure*}

Instead of using a same learning rate for all layers in the network, we utilize layer-wise learning rates in the training phase. Specifically, learning rates for the convolutional layer in feature fusion block, two Type-1 blocks, two Type-2 blocks and the fully-connected layer are set to 0.01, 0.001, 0.0001 and 0.0001 respectively. As the network is optimized for image steganalysis, feature maps of cover images and their stegos are more discrimnative in later layers than they are in initial layers. Therefore, the layer-wise learning rate strategy can make different layers in network be optimized in a same speed. To make the HPFs adaptive to the network, we also let them be updated in the learning stage and set its learning rate to be a small value, i.e. $5 \times 10^{-6}$. During the training, all learning rates of our model decay with an exponential factor, 0.985:
\begin{equation}
  \alpha_{i}(t) = \alpha_{i} \cdot 0.985^{t}
\end{equation}
where $t$ represents the epoch number, $\alpha_{i}$ denotes the learning rate for convolutional/fully-connected layers, i.e. $\alpha_{i} \in \{ 5\times 10^{-6}, 0.01, 0.001, 0.0001, 0.0001 \}$.

It is usually hard for deep learning based methods to directly learn discriminative features between cover images and stego images when the payload is low [25,32]. Compared with steganography at high payloads, modifications introduced by steganographic embedding at low payloads are too weak and almost all of them are located in regions with highly varied intensities. In this case, deep models are difficult to discriminate cover images and their stegos since high frequency components of cover images may swamp the existence of embedded message, making image steganalysis more challenging. To handle the difficulty, we use curriculum learning [48] to detect steganopgraphic algorithms at low payloads. Specifically, the CIS-Net trained for a lower payload steganalysis, e.g. 0.4 bpp, are refined on the network trained at a higher payload, e.g. 0.5 bpp. The advantage of curriculum learning is that attentional features learned by higher payloads can guide/regularize the search of locations modified by steganographic algorithms at low payloads, which make the task much easier. To avoid training samples are reused for testing at different payloads, we force that cover images and their stegos used for network training/testing at a lower payload are same to those used for network training/testing at a higher payload.

\subsection{Ablation Study}
This section conducts ablation study to the proposed network. We examine the behavior of proposed CIS-Net at different configuration of STL and SPL. The S-UNIWARD steganography at 0.4 bpp is used for performance validation in following experiments.

\textit{Truncation threshold T in STL}. In this subsection, we evaluate the performance of our CIS-Net at different truncation thresholds $T$. To demonstrate the effectiveness of proposed STL, the network equipped with Bi-valued Truncation Layer (BTL) is also compared in the experiment.

Fig.5(a) shows detection error rates of the network with STL and BTL at seven different thresholds, i.e. $T \in [1,3,5,7,11,+\infty]$, where $+\infty$ means that no truncation is applied in the model. Results in the figure indicate that the model with STL is systematically better than the model with BTL at different $T$, which demonstrates the effectiveness of proposed data truncation method for image steganalysis. For small $T$ value, e.g. $T=1$, the detection error rate of the CIS-Net is very high because excessive truncation to feature map elements makes discriminative features between cover images and stego images lost. For large $T$ values, e.g. $T>5$, the network's performance degrades due to negative influence caused by large elements in cover image content. Therefore, to set an appropriate truncation threshold $T$ is important for CNN based image steganalysis.

\textit{Power factor $(\gamma_{1}, \gamma_{2})$ in SPL}. This subsection studies how the power of sublinear function in a SPL affects the performance of CIS-Net for detecting steganography. Both $\gamma_{1}$ and $\gamma_{2}$ in proposed SPL are selected from a given set $[0.6, 0.7, 0.8, 0.9, 1.0]$. When both $\gamma_{1}$ and $\gamma_{2}$ are equal to 1.0, the SPL becomes a normal average pooling. In the experiment, two Type-2 blocks use same setting of $\gamma_{1}$ and $\gamma_{2}$. Fig.5(b) shows detection error rates when different configurations of $\gamma_{1}$ and $\gamma_{2}$ are set to the CIS-Net. From the figure, we can observe that detection error rates are high when $\gamma_{1}$ and $\gamma_{2}$ are low values. The reason is that low power factors not only reduce cover image content but also remove the embedded message, making classification of cover images and stego images difficult. The best performance ($14.62\%$ detection error rate) is obtained when two power factors are around 1.0, i.e. $(\gamma_{1},\gamma_{2}) = (1.0, 0.9)$. The result indicates that a slight sublinear suppression to feature map after average pooling is beneficial for image steganalysis. In the following experiments, we use this setting to train the model for different steganographic algorithms at five payloads.

\subsection{Performance Comparisons with Prior Arts}
In this section, we conduct experiments to demonstrate the effectiveness of proposed CIS-Net for image steganalysis. Two kinds of methods, i.e. hand-crafted feature based methods and deep learning based methods, are compared in the experiment. For hand-crafted feature based methods, we compare performances of the proposed CIS-Net with the classic SRM steganalysis [11] and its selection channel version, the maxSRMd2 steganalysis [12]. For deep learning based methods, four state of arts CNN models including Xu-Net [28], Ye-Net [32], Share Normalization Network (SN-Net) [25] and ReST-Net [29] are selected for performance comparison. For ReST-Net, we both compare the model with SRM high pass filers. Three steganographic algorithms, WOW, S-UNIWARD and HILL, at five different payloads [0.1, 0.2, 0.3, 0.4, 0.5] are evaluated. In the experiment, 5,000 cover images and their corresponding stego images are randomly selected to train the model, the rest 5,000 cover images and their stegos are used for testing. To make results reliable, all reported detection error rates of the proposed model are results of the average performance of 5 times running.

TABLE II and Fig.6 show detection error rates of the SRM, maxSRMd2 and the proposed CIS-Net on three steganographic algorithms at five payloads. Additionally, ROC curves of three methods at payload 0.4 bpp and 0.2 bpp are provided in Fig.7. Compared with the SRM steganalysis, our model obtains significant performance gain on different steganographic algorithms. Furthermore, the proposed CIS-Net outperforms the maxSRMd2 steganalysis even no information of selection channel is provided. An interesting phenomenon in TABLE II is that, compared to SRM steganalysis, CIS-Net's performance gain over maxSRMd2 steganalysis decreases as the payload decreases. There are two main reasons for such phenomenon. One is that secret messages are embedded at complex/cluttered regions in a cover image when the payload is low. After highpass filtering, those regions in cover images are statistically similar to stego images. The other is that the maxSRMd2 steganalysis only extracts features at positions that secret messages are exactly embedded since it is provided with embedding probability map. However, CIS-Net only has a rough estimation to embedding positions provided by the network trained at high payloads.

Despite the traditional steganalytic methods, we also compare the proposed method with four deep CNN models on S-UNIWARD and HILL steganography at five payloads. For Xu-network and ReST network, we report their performances according to Li's paper [29]. For Ye-network, it is originally a CNN model optimized for $256 \times 256$ input images. Li in [29] has implemented a version of Ye-net which can detect $512 \times 512$ spatial images. Here, we use such result for performance comparison. For SN-Network, the performances in [25] are used for reporting. Recently, several research papers [34][51] used both BOSS [44] and BOWS [52] as the training set and obtain promising performance. However, these networks are only optimized for downsampled images ($256 \times 256$) and use more training data in model learning. Such setting is greatly different from our case. Therefore, these networks are not compared in our experiments. Results in TABLE III show that the proposed CIS-Net outperforms Xu-network, Ye-network, SN-Network and ReST-Net (SRM) on all configurations.

{\setlength{\abovecaptionskip}{2pt}
 \setlength{\belowcaptionskip}{-2pt}
\begin{table}[t]
  \centering
  \renewcommand\arraystretch{1.2}
  \caption{Detection error rates of proposed CIS-Net and ReST-Net (ensemble) on S-UNIWARD and HILL at five different payloads. }
  \resizebox{9.0cm}{!} {
  \begin{tabular}{| c | c | c | c | c | c |}
   \hline
  \textbf{S-UNIWARD} & 0.1 bpp & 0.2 bpp & 0.3 bpp & 0.4 bpp & 0.5 bpp \\ \hline
  ReST-Net (ensemble) & \textbf{34.33\%} & 28.65\% & 21.22\% & \textbf{14.56\%} & 12.07\% \\ \hline
  Proposed CIS-Net & 35.28\% & \textbf{26.21\%} & \textbf{19.64\%} & 14.62\% & \textbf{10.73\%} \\ \hline \hline
  \textbf{HILL} & 0.1 bpp & 0.2 bpp & 0.3 bpp & 0.4 bpp & 0.5 bpp \\ \hline
  ReST-Net (ensemble) & 37.62\% & 29.36\% & 23.26\% & 18.34\% & 15.46\% \\ \hline
  Proposed CIS-Net & \textbf{36.82\%} & \textbf{28.83\%} & \textbf{22.67\%} & \textbf{18.10\%} & \textbf{14.78\%} \\ \hline
  \end{tabular}
  }
\end{table}}

Li in [29] boosted the performance of ReST-Net via ensembling three networks (ReST-Net ensemble), in which each network is equipped with three different highpass filters, i.e. SRM, Gabor filters and the max-min nonlinear filters, for feature extraction. In our experiment, we also compare the proposed network with ReST-Net (ensemble) on S-UNIWARD and HILL. Table IV shows that our CIS-Net outperforms ReST-Net (ensemble) in most of configurations even though it is augmented by model ensemble.

{\setlength{\abovecaptionskip}{2pt}
 \setlength{\belowcaptionskip}{-2pt}
\begin{table}[t]
  \centering
  \renewcommand\arraystretch{1.2}
  \caption{Detection error rates of proposed model with and without augmentation on three steganographic algorithms at payload 0.4 bpp. }
  \resizebox{8.4cm}{!} {
  \begin{tabular}{| c | c | c | c |}
   \hline
  \textbf{Method} & WOW & S-UNIWARD & HILL \\ \hline
  No augmentation & 12.13\% & 14.62\% & 18.10\% \\ \hline
  Augmentation  & \textbf{11.56\%} & \textbf{13.95\%} & \textbf{17.62\%} \\ \hline
  \end{tabular}
  }
\end{table}}

Recent researches in deep learning show that data augmentation is important for the performance improvement of various CNN models [49-50]. In this experiment, we use data augmentation method to decrease the detection error rates of CIS-Net for steganographic algorithms. Same to the setting in [25], we randomly split 10,000 BOSSbase samples into 5,000 training images and 5,000 testing images. For training images, we rotate them with 90 degree, 180 degree, and 270 degree along counter clockwise direction, which generates a new training set with 20,000 samples. Then, three steganographic algorithms embed secret messages into the augmented training set and the test set.  The proposed network is trained on this new training set with 20,000 covers/stegos and finally validated on the test set with 5,000 covers/stegos. To make experiment simple, we only demonstrate the performance of proposed CIS-Net at payload 0.4 bpp. Detection error rates in TABLE V indicate that data augmentation can improve the performance of proposed CIS-Net on different algorithms.

\subsection{Attentional Map Extraction for CIS-Net }
In [38], Zhou et al. showed that an image classification CNN exposes implicit attention of the model on an image. Such ability of CNN models can be used to localize most discriminative regions contributing to image classification. For steganalytic CNN models, similar idea was also reported in [32] that the network can defeat the selection-channel-aware maxSRMd2 steganalytic algorithm, demonstrating that they are able to implicitly learn the distribution of selection channel for a specific embedding scheme. In this experiment, we aim to draw attentional features learned by the proposed CIS-Net for given images. For image steganalysis, such attentional feature is actually the estimation to the embedding probability map of the steganographic algorithm. The motivation is to understand whether a well trained CNN can indeed extract the embedding probability map implicitly even no such information is provided. Additionally, we also want to analyze the difference between the estimated embedding probability map and true embedding probability map, which may reveal limitations of CNN models for image steganalysis.

The Class Activation Mapping (CAM) [38] is an effective method to extract attentional maps learned by a CNN model. Specifically, it computes a weighted sum of CNN's last feature maps as follows:
\begin{equation}
  M_{c}(x,y) = \sum_{k} w_{k}^{c}f_{k}(x,y)
\end{equation}
where $f_{k}(x,y)$ represents the feature map of unit $k$ in the last convolutional layer at spatial location $(x,y)$, and $w_{k}^{c}$ is the learned weight in the fully connect layer corresponding to class $c$ for unit $k$. CAM highlights discriminative visual patterns for class $c$ represented by $f_{k}(x,y)$ using the weight matrix $w_{k}^{c}$. For adaptive steganography, the discrimination between cover images and stego images mainly comes from noisy/cluttered regions in which secret messages are mostly embedded. Therefore, attentional feature map extracted by CAM for image steganalysis is an estimation to the embedding probability map.

Following the idea of CAM method, we compute a weighted sum of CIS-Net's feature maps of the global average pooling in second SPL to obtain attentional maps. To make the size of attentional maps comparable to input images, we simply resize them from $64 \times 64$ to $512 \times 512$ with ``imresize'' in Matlab. In the experiment, we randomly select several cover images from BOSSbase 1.01 and also provide their ground truth embedding probability maps for S-UNIWARD steganography at 0.4 bpp. Fig.(8) shows five cover images, their ground truth embedding probability maps and the attentional maps calculated by CAM respectively. From the figure, we can easily observed that attentional maps extracted by CAM are visually similar to ground truth embedding probability maps. The observation indicates that our proposed CIS-Net can implicitly estimate positions of embedded messages in case that no selected channel information is provided. In addition, we also compare the differences between CAM attentional maps of three steganographic algorithms in Fig.9. Compare to the S-UNIWARD and HILL steganography, the attentional map of WOW is almost equal to zero at the region without message embedding. This demonstrates that CIS-Net optimized for WOW only extract discriminative features at message embedding regions, thus the detection error rate should be low. However, for the attentional map of HILL, it is still activated at no message embedding regions. These noisy activations are very harmful for discriminative features extraction between cover images and their stegos, thus make image steganalysis difficult. Therefore, the detection error rate should be high. Such analysis from extracted attentional map of three steganographic algorithms is consistent with results reported in TABLE II, indicating that the quality of CAM attentional map is consistent with the performance of CNN model for steganographic algorithms. The reason why attentional maps of different steganographic algorithms demonstrate different visual qualities is that the embedding method of WOW and S-UNIWARD make all secret messages be crowded in complex regions, while HILL use ``spreading strategy'' to make messages be distributed around complex regions. This strategy not only decreases the embedding intensity in a local region but also spreads secret messages into high frequency component of cover images. In this case, a CNN model is hard to classify cover images and their stegos since the embedding message signals and high frequency cover image components are mixed together.

\section{Conclusion}
In this paper, we propose a novel CNN model called CIS-Net to detect adaptive steganography in spatial domain. Two new layers, i.e. single-valued truncation layer and sublinear pooling layer, are designed to suppress cover image content. The single-valued truncation layer uses a same truncation threshold to reduce the variance introduced by the truncated data, while the sublinear pooling layer adaptively suppresses large elements of cover image content and aggregate weak embedded message signal with average pooling. Compared with previous data truncation and feature pooling, the proposed two layers can accelerate the learning and improve the generalization ability of the CNN model. Additionally, we use class activation map method to demonstrate that the proposed CIS-Net can learn the embedding probability map of steganographic algorithms when no selection channel information is provided. The result shows that CNN models have the ability to estimate the message embedding positions implicitly. In future works, we would extend our methods to compressed domain images.

\section{Appendix}
In this appendix, we prove that the proposed STL can reduce the variance of traditional data truncation method. For Eq.(4), we expand it as following equations:
\begin{equation}
 \begin{aligned}
   &\sigma^{2}_{s} = 2\int_{T}^{\infty}\frac{(T-\mu_{s})^{2}}{Z}e^{-\left|\frac{x}{s}\right|^{\alpha}}dx + \int_{-T}^{T}\frac{(x-\mu_{s})^2}{Z}e^{-\left|\frac{x}{s}\right|^{\alpha}}dx \\
   &= 2\int_{T}^{\infty}\frac{T^{2}}{Z}e^{-\left|\frac{x}{s}\right|^{\alpha}}dx + \int_{-T}^{T}\frac{x^2}{Z}e^{-\left|\frac{x}{s}\right|^{\alpha}}dx \\
   &+  \int_{-T}^{T}\frac{\mu_{s}^2}{Z}e^{-\left|\frac{x}{s}\right|^{\alpha}}dx + 2\int_{T}^{\infty}\frac{\mu_{s}^{2}}{Z}e^{-\left|\frac{x}{s}\right|^{\alpha}}dx \\
   &- 2\mu_{s}\int_{-T}^{T}\frac{x}{Z}e^{-\left|\frac{x}{s}\right|^{\alpha}}dx - 4\mu_{s}\int_{T}^{\infty}\frac{T}{Z}e^{-\left|\frac{x}{s}\right|^{\alpha}}dx
 \end{aligned}
\end{equation}
for the third line of Eq.(19), it can be written as:
\begin{equation}
 \begin{aligned}
  &\int_{-T}^{T}\frac{\mu_{s}^2}{Z}e^{-\left|\frac{x}{s}\right|^{\alpha}}dx + 2\int_{T}^{\infty}\frac{\mu_{s}^{2}}{Z}e^{-\left|\frac{x}{s}\right|^{\alpha}}dx \\
  &=\int_{-\infty}^{-T}\frac{\mu_{s}^{2}}{Z}e^{-\left|\frac{x}{s}\right|^{\alpha}}dx + \int_{-T}^{T}\frac{\mu_{s}^2}{Z}e^{-\left|\frac{x}{s}\right|^{\alpha}}dx + \int_{T}^{\infty}\frac{\mu_{s}^{2}}{Z}e^{-\left|\frac{x}{s}\right|^{\alpha}}dx \\
  &=\mu_{s}^{2}\int_{-\infty}^{\infty}\frac{1}{Z}e^{-\left|\frac{x}{s}\right|^{\alpha}}dx = \mu_{s}^{2}
 \end{aligned}
\end{equation}
since $p(x)$ is a symmetric function, the following integral is equal to zero:
\begin{equation}
  2\mu_{s}\int_{-T}^{T}\frac{x}{Z}e^{-\left|\frac{x}{s}\right|^{\alpha}}dx = 0
\end{equation}
based on Eq.(7), we obtain:
\begin{equation}
  4\mu_{s}\int_{T}^{\infty}\frac{T}{Z}e^{-\left|\frac{x}{s}\right|^{\alpha}}dx=2\mu_{s} \cdot 2\int_{T}^{\infty}\frac{T}{Z}e^{-\left|\frac{x}{s}\right|^{\alpha}}dx = 2\mu_{s}^{2}
\end{equation}
Combining Eq.(20), Eq.(21) and Eq.(22), $\sigma^{2}_{s}$ can be written as Eq.(9).

\ifCLASSOPTIONcaptionsoff
  \newpage
\fi


\begin{thebibliography}{1}

\bibitem{IEEEhowto:kopka}
J. Mielikainen, ``LSB matching revisited,'' \emph{IEEE Signal Processing Letters}, 13(5):285-287, 2006.

\bibitem{IEEEhowto:kopka}
X. Zhang and S. Wang, ``Efficient steganographic embedding by exploiting modification direction,'', \emph{IEEE Communications Letters}, 10(11):781-783, 2006.

\bibitem{IEEEhowto:kopka}
W. Luo, F. Huang, and Jiwu Huang, ``Edge adaptive image steganography based on LSB matching revisited,'' \emph{IEEE Transactions on Information Forensics and Security}, 5(2):201-214, 2010.

\bibitem{IEEEhowto:kopka}
T. Filler and J. Fridrich, ``Gibbs construction in steganography,'' \emph{IEEE Transactions on Information Forensics and Security}, 5(4):705-720,2010.

\bibitem{IEEEhowto:kopka}
V. Holub and J. Fridrich, ``Designing steganographic distortion using directional filters,'' \emph{IEEE Workshop on Information Forensic and Security}, 2012.

\bibitem{IEEEhowto:kopka}
V. Holub, J. Fridrich, and T. Denemark, ``Universal distortion function for steganography in an arbitrary domain,'' \emph{EURASIP Journal on Information Security}, 1(1):1-13, 2014.

\bibitem{IEEEhowto:kopka}
B. Li, M. Wang, J. Huang, and X. Li, ``A new cost function for spatial image steganography,'' \emph{IEEE International Conference on Image Processing}, pp.4206-4210, 2014.

\bibitem{IEEEhowto:kopka}
T. Denemark and J. Fridrich, ``Improving steganographic security by synchronizing the selection channel,'' \emph{Proceedings of 3rd ACM Workshop on Information Hiding and Multimedia Security}, 2015.

\bibitem{IEEEhowto:kopka}
S. Lyu and H. Farid, ``Detecting hidden messages using higher-order statistics and support vector machines,'' \emph{International Workshop on Information Hiding}, 2002.

\bibitem{IEEEhowto:kopka}
J. Fridrich, ''Feature-based steganalysis for JPEG images and its implications for future design of steganographic schemes,'' \emph{International Workshop on Information Hiding}, pp.67-81, 2004.

\bibitem{IEEEhowto:kopka}
T. Pevny, P. Bas, and J. Fridrich, ''Steganalysis by subtractive pixel adjacency matrix,'' \emph{IEEE Transactions on Information Forensics and Security}, 5(2):215-224, 2010.

\bibitem{IEEEhowto:kopka}
J. Fridrich and J. Kodovsky, ``Rich models for steganalysis of digital images,'' \emph{IEEE Transactions on Information Forensics and Security}, 7(3):868-882, 2012.

\bibitem{IEEEhowto:kopka}
T. Denemark, V. Sedighi, V. Holub, R. Cogranne, and J. Fridrich, ``Selection-channel-aware rich model for steganalysis of digital images,'' \emph{IEEE Workshop on Information Forensic and Security}, 2014.

\bibitem{IEEEhowto:kopka}
V. Holub and J. Fridrich, ``Random projections of residuals for digital image steganalysis,'' \emph{IEEE Transactions on Information Forensics and Security}, 8(12):1996-2006, 2013.

\bibitem{IEEEhowto:kopka}
H. Yin, W. Hui, H. Li, C. Lin, and W. Zhu, ``A novel large-scale digital forensics service platform for internet videos,'' \emph{IEEE Transactions on Multimedia}, 14(1):178-186, 2012.

\bibitem{IEEEhowto:kopka}
H. Zhou, K. Chen, W. Zhang, C. Qin, and N. Yu, ``Feature-preserving tensor voting model for mesh steganalysis,'' \emph{IEEE Transactions on Visualization and Computer Graphics}, DOI:10.1109/TVCG.2019.2929041, 2019.

\bibitem{IEEEhowto:kopka}
T. Filler, J. Judas, and J. Fridrich, ``Minimizing additive distortion in steganography using syndrome-trellis codes,'' \emph{IEEE Transactions on Information Forensics and Security}, 6(3):920-935, 2011.

\bibitem{IEEEhowto:kopka}
T. Pevny and J. Fridrich, ``Merging Markov and DCT features for multi-class JPEG steganalysis,'', \emph{Proceedings of SPIE Electronic Imaging}, 2007.

\bibitem{IEEEhowto:kopka}
A. D. Ker, P. Bas, R. Bohme, R. Cogranne, S. Craver, T. Filler, J. Fridrich, and T. Pevny, ``Moving steganography and steganalysis from the laboratory into the real world,'' \emph{Proceedings of the first ACM workshop on Information Hiding and Multimedia Security}, pp.45-58, 2013.

\bibitem{IEEEhowto:kopka}
K. He, X. Zhang, S. Ren, and J. Sun, ``Deep residual learning for image recognition,'' \emph{IEEE Conference on Computer Vision and Pattern Recognition}, 2016.

\bibitem{IEEEhowto:kopka}
I, Goodfellow, J. Pouget-Abadie, M. Mirza, B. Xu, D. Warde-Farley, S. Ozair, A. Courville, and Y. Bengio, ``Generative adversarial nets,'' \emph{Advances in Neural Processing Systems}, pp.2672-2680, 2014.

\bibitem{IEEEhowto:kopka}
C. Dong, C. Loy, K. He, and X. Tang, ``Image super-resolution using deep convolutional networks,'' \emph{IEEE Transactions on Pattern Analysis and Machine Intelligence}, 38(2):295-307, 2015.

\bibitem{IEEEhowto:kopka}
V. Badrinarayanan, A. Kendall, and R. Cipolla, ``SegNet: A deep convolutional encoder-decoder architecture for image segmentation,'' \emph{IEEE Transactions on Pattern Analysis and Machine Intelligence}, 39(12):2481-2495, 2017.

\bibitem{IEEEhowto:kopka}
K. Zhang, W. Zuo, Y. Chen, D. Meng, and L. Zhang, ``Beyond a gaussian denoiser: Residual learning of deep cnn for image denoising,'' \emph{IEEE Transactions on Image Processing}, 26(7):3142-3155, 2017.

\bibitem{IEEEhowto:kopka}
S. Wu, S. Zhong, and Y. Liu, ``A novel convolutional neural network for image steganalysis with shared normalization,'' \emph{IEEE Transactions on Multimedia}, DOI: 10.1109/TMM.2019.2920605, 2019.

\bibitem{IEEEhowto:kopka}
S. Tan and B. Li, ``Stacked convolutional auto-encoders for steganalysis of digital images,'' \emph{Asia-Pacific Signal and Information Processing Association, 2014 Annual Summit and Conference (APSIPA)}, 2014, pp.1-4.

\bibitem{IEEEhowto:kopka}
Y. Qian, J. Dong, W. Wang, and T. Tan, ``Deep learning for steganalysis via convolutional neural networks,'' \emph{SPIE Media Watermarking, Security, and Forensics}, vol. 9409, 2015.

\bibitem{IEEEhowto:kopka}
G. Xu, H. Z. Wu, and Y. Q. Shi, ``Structural design of convolutional neural networks for steganalysis,'' \emph{IEEE Signal Processing Letters}, 23(5):708-712, 2016.

\bibitem{IEEEhowto:kopka}
B. Li, W. Wei, A. Ferreira, and S. Tan, ``ReST-Net: Diverse activation modules and parallel subnets-based CNN for spatial image steganalysis,'' \emph{IEEE Signal Processing Letters}, 25(5):650-654, 2018.

\bibitem{IEEEhowto:kopka}
S. Wu, S. Zhong, and Y. Liu, ``Deep residual learning for image steganalysis,'' \emph{Multimedia Tools and Applications}, pp. 1-17, 2017.

\bibitem{IEEEhowto:kopka}
S. Wu, S. Zhong, and Y. Liu,``Residual convolution network based steganalysis with adaptive content suppression,'' \emph{IEEE International Conference on Multimedia and Expo (ICME)}, 2017.

\bibitem{IEEEhowto:kopka}
J. Ye, J. Ni, and Y. Yi, ``Deep learning hierarchical representations for image steganalysis,''  \emph{IEEE Transactions on Information Forensics and Security}, 12(11):2545-2557, 2017.

\bibitem{IEEEhowto:kopka}
W. Wang, J. Dong, Y. Qian, and T. Tan, ``Deep steganalysis: End-to-end learning with supervisory information beyond class labels,'' \emph{arXiv:1806.10443v1}, 2018.

\bibitem{IEEEhowto:kopka}
M. Boroumand, M. Chen, and J. Fridrich, ``Deep residual network for steganalysis of digital images,'' \emph{IEEE Transactions on Information Forensics and Security}, 14(5):1181-1193, 2018.

\bibitem{IEEEhowto:kopka}
M. Chen, V. Sedighi, M. Boroumand, and J. Fridrich, ``JPEG-phase-aware convolutional neural network for steganalysis of JPEG images,'' \emph{Proceedings of the 5th ACM Workshop on Information Hiding and Multimedia Security}, pp.75-84, 2017.

\bibitem{IEEEhowto:kopka}
J. Zeng, S. Tan, B. Li, and J. Huang, ``Large-scale JPEG image steganalysis using hybrid deep-learning framework,'' \emph{IEEE Transactions on Information Forensics and Security}, 13(5):1200-1214, 2017.

\bibitem{IEEEhowto:kopka}
G. Xu, ``Deep convolutional neural network to detect J-UNIWARD,'' \emph{Proceedings of the 5th ACM Workshop on Information Hiding and Multimedia Security}, pp.67-73, 2017.

\bibitem{IEEEhowto:kopka}
B. Zhou, A. Khosla, A. Lapedriza, A. Oliva, and A. Torralba, ``Learning deep features for discriminative localization,'' \emph{IEEE Conference on Computer Vision and Pattern Recognition}, 2016.

\bibitem{IEEEhowto:kopka}
K. Simonyan and A. Zisserman, ``Very deep convolutional networks for large-scale image recognition,'' \emph{International Conference on Learning Representation}, 2015.

\bibitem{IEEEhowto:kopka}
F. Yu and V. Koltun, ``Multi-scale context aggregation by dilated convolutions,`` \emph{International Conference on Learning Representation}, 2016.

\bibitem{IEEEhowto:kopka}
J. Huang, ``Statistics of natural images and models,'' \emph{PhD Thesis, Brown University}, 2000.

\bibitem{IEEEhowto:kopka}
A. Srivastava, A. B. Lee, E. P. Simoncelli, and S-C. Zhu, ``On advances in statistical modeling of natural images,'' \emph{Journal of Mathematical Imaging and Vision}, 18(1):17-33, 2003.

\bibitem{IEEEhowto:kopka}
M. Simon, Y. Gao, T. Darrell, J. Denzler, and E. Rodner, ``Generalized orderless pooling performs implicit salient matching,'' \emph{IEEE International Conference on Computer Vision}, 2017.

\bibitem{IEEEhowto:kopka}
P. Bas, T. Filler, and T. Pevny, ``Break our steganographic system: the ins and outs of organizing BOSS,'' \emph{International Workshop on Information Hiding}, pp.59-70, 2011.

\bibitem{IEEEhowto:kopka}
L. Pibre, J. Pasquet, J. Pasquet, D. Ienco, D. Ienco, and M. Chaumont, ``Deep learning is a good steganalysis tool when embedding key is reused for different images, even if there is a cover sourcemismatch,'' \emph{Media Watermarking, Security, and Forensics, Part of IS\&T International Symposium on Electronic Imaging}, 2016.

\bibitem{IEEEhowto:kopka}
K. He, X. Zhang, S. Ren, and J. Sun, ``Delving deep into rectifiers: surpassing human-level performance on imageNet classification,'' \emph{IEEE International Conference on Computer Vision}, 2015.

\bibitem{IEEEhowto:kopka}
D. P. Kingma and J. L. Ba, ``Adam: A method for stochastic optimization,'' \emph{International Conference on Learning Representation}, 2015.

\bibitem{IEEEhowto:kopka}
Y. Bengio, J. Louradour, R. Collobert, and J. Weston, ``Curriculum learning,'' \emph{International Conference on Machine Learning}, 2009.

\bibitem{IEEEhowto:kopka}
L. Perez and J. Wang, ``The effectiveness of data augmentation in image classification using deep learning,'' \emph{arXiv:1712.04621}, 2017.

\bibitem{IEEEhowto:kopka}
Y. Xu, R. Jia, L. Mou, G. Li, Y. Chen, Y. Lu, and Z. Jin, ``Improved relation classification by deep recurrent neural networks with data augmentation,'' \emph{arXiv:1601.03651v2}, 2016.

\bibitem{IEEEhowto:kopka}
R. Zhang, F. Zhu, J. Liu, G. Liu, ``Depth-wise separable convolutions and multi-level pooling for an efficient spatial CNN-based steganalysis,'' \emph{IEEE Transactions on Information Forensics and Security}, DOI: 10.1109/TIFS.2019.2936913, 2019.

\bibitem{IEEEhowto:kopka}
P. Bas and T. Furon. ``Breaking Our Watermarking System (BOWS)''. Available: http://bows2.gipsa-lab.inpg.fr. 2007.

\end{thebibliography}
\end{document}